\documentclass{ws-ijmpcs}

\usepackage{graphicx}
\usepackage{theorem}
\usepackage{color}
\usepackage{bm}
\usepackage{dcolumn,epsfig}
\usepackage{wasysym}

\newcommand{\bit}{\begin{itemize}}
\newcommand{\eit}{\end{itemize}}

\newcommand{\der}[2]{\frac{\partial #1}{\partial #2}}

\newcommand{\w}[1]{\bm{#1}}
\newcommand{\be}{\begin{equation}}
\newcommand{\ee}{\end{equation}}
\newcommand{\bea}{\begin{eqnarray}}
\newcommand{\eea}{\end{eqnarray}}
\newcommand{\nn}{\nonumber}
\newcommand{\noi}{\noindent}
\newcommand{\Hor}{{\mathcal H}}

\newcommand{\Sp}{{\mathcal S}}

\newcommand{\el}{\w{\ell}}

\newcommand{\tDS}{{}^{2}\!\tilde D}

\newcommand{\tD}{\tilde D}

\font\tenscr=rsfs10 scaled1100
\font\sevenscr=rsfs7 % scaled \magstep1
\font\fivescr=rsfs5 % scaled \magstep1
\skewchar\tenscr='177
\skewchar\sevenscr='177
\skewchar\fivescr='177
\newfam\scrfam
\textfont\scrfam=\tenscr
\scriptfont\scrfam=\sevenscr
\scriptscriptfont\scrfam=\fivescr

\def\scri{{\fam\scrfam I}}

\def\scre{{\fam\scrfam E}}
\def\scrb{{\fam\scrfam B}}

\begin{document}

\markboth{J.L. JARAMILLO}
{
%AN INTRODUCTION TO SOME ASPECTS OF LOCAL BLACK HOLE HORIZONS
LOCAL BLACK HOLE HORIZONS IN THE 3+1 APPROACH
TO GENERAL RELATIVITY
}

\catchline{}{}{}{}{}
%
%%%%%%%%%%%%%%%%%%%%%%%%%%%%%%%%%%%%%%%%%%%%%%%%%%%%%%%%%%%%%%%%%%%%

\title{
%AN INTRODUCTION TO SOME ASPECTS OF LOCAL BLACK HOLE HORIZONS: A
%3+1 PERSPECTIVE
AN INTRODUCTION TO LOCAL BLACK HOLE HORIZONS IN THE 3+1 APPROACH
TO GENERAL RELATIVITY
}

\author{JOS\'E LUIS JARAMILLO
%\footnote{
%Typeset names in 8 pt roman, uppercase. Use the footnote to indicate the
%present or permanent address of the author.}
}

\address{
Max-Planck-Institut f{\"u}r Gravitationsphysik, Albert Einstein
Institut\\ 
Am M{\"u}hlenberg 1, Golm D-14476, Germany
%University Department, University Name, Address\\
%City, State ZIP/Zone,
%Country
%\footnote{State completely without abbreviations, the
%affiliation and mailing address, including country. Typeset in 8 pt
%italic.}
\\
Jose-Luis.Jaramillo@aei.mpg.de}

%\author{SECOND AUTHOR}

%\address{Group, Laboratory, Address\\
%City, State ZIP/Zone, Country\\
%second\_author@domain\_name}

\maketitle

\begin{history}
\received{Day Month Year}
\revised{Day Month Year}
\end{history}

\begin{abstract}

We present an introduction to dynamical trapping horizons 
as quasi-local models for black hole horizons, 
from the perspective of an Initial Value Problem approach
to the construction of generic black hole spacetimes.
We focus on the geometric and structural properties 
of these horizons aiming, as a main application, at the numerical evolution and
analysis of black hole spacetimes in astrophysical scenarios.
In this setting, we discuss
their dual role as an {\em a priori} ingredient in 
certain formulations of Einstein equations and as an
{\em a posteriori} tool for the diagnosis of dynamical black hole spacetimes.
Complementary to the  first-principles discussion of quasi-local horizon
physics, we place an emphasis on the {\em rigidity} properties of these hypersurfaces
and their role as privileged geometric probes into near-horizon 
strong-field spacetime dynamics.

%study of gravity in the strong-field regime, where the lack of {\em rigid structures}
%(e.g. symmetries, background spacetime...) represents a generic problem. 

\end{abstract}

\ccode{PACS numbers: 04.70.Bw, 	04.20.Ex, 04.25.dg}

\section{Black holes: global vs. (quasi-)local approaches}
\label{s:global_local}

\subsection{{\em Establishment's picture} of the gravitational collapse}
Our discussion is framed in the  problem
of gravitational collapse in General Relativity. The current
understanding is summarized in what one could 
call the {\em establishment's picture} of gravitational collapse\cite{Pen73},
a heuristic chain of results and conjectures:
\begin{enumerate}
\item {\em Singularity Theorems}: if gravity is able to make all light rays 
locally {\em converge}  (namely,
if trapped surfaces exist), then
a spacetime singularity forms\cite{Pen65b,Hawki67,HawPen70,HawEll73}.
\item {\em (Weak) Cosmic Censorship} (Conjecture): 
in order to preserve predictability, the formed singularity is not visible for a
distant observer\cite{Pen69}.
\item {\em Black hole spacetimes stability} (Conjecture):
General Relativity gravitational dynamics drives eventually the black hole
spacetime to a stationary state.
\item {\em Black Hole uniqueness theorem}: the
final state is a Kerr black hole spacetime\cite{Heusl98}.
\end{enumerate} 
Light bending is a manifestation of spacetime curvature and
black holes constitute a dramatic extreme case of this.
The standard picture of gravitational
collapse above suggests two (complementary) approaches to the
 characterization of  black holes:
\begin{itemize}
\item[a)] {\em Global approach}: (weak) cosmic censorship suggests 
black holes as {\em no-escape} regions 
not extending to infinity. Its boundary defines the event horizon $\scre$. 
\item[b)] {\em Quasi-local approach}: singularity theorems suggest
the characterization of a black hole as a spacetime {\em trapped region}
where all light rays locally converge. 
\end{itemize}
The {\em establishment's picture} of gravitational collapse
depicts an intrinsically dynamical scenario. Hence, a systematic
methodology to the study of dynamical spacetimes is needed.
We adopt an {\em Initial (Boundary) Value Problem} approach,
that offers a systematic avenue to the {\em qualitative} and {\em quantitative}
aspects of generic spacetimes.

\subsection{The Black Hole region and the Event Horizon}
\label{s:Event_Horizon}
The traditional\cite{HawEll73} approach to black holes
involves global spacetime concepts, in particular a 
good control of the notion of infinity.
Given a (strongly asymptotically predictable) spacetime ${\cal M}$,
the {\em black hole region} $\scrb$ is defined as 
$\scrb = {\cal M} - J^-(\scri^+)$, 
where $J^-(\scri^+)$ is the causal past of future null infinity $\scri^+$.
That is, $\scrb$ is the spacetime region that cannot communicate
with $\scri^+$.

We are particularly interested in characterizing 
a notion of boundary {\em surface} of black holes.
In this global context this is provided  by the event horizon $\scre$,
defined as the boundary of $\scrb$, that is 
$\scre = {\partial{J}}^-(\scri^+)\cap {\cal M}$. Interesting geometric and
physical properties of the event horizon are: i) 
$\scre$ is null hypersurface in ${\cal M}$; 
ii) it satisfies an {\em Area Theorem}\cite{Hawki71,Hawki72},
so that the area of spatial sections ${\cal S}$ of  $\scre$ does not decrease
in the evolution; and, beyond that,  iii) a set of {\em black hole mechanics} laws
are fulfilled\cite{BarCarHaw73}. 

However, the global aspects of the event horizon 
also bring difficulties: a) it is a {\em teleological} concept, i.e. 
the  knowledge of the
full (future) spacetime is needed in order to locate  $\scre$, and
b) the black hole region and the event horizon can enter into flat 
spacetime regions.
In sum, the notion of event horizon is a too global one: 
it does not fit properly into the adopted  Initial Value Problem approach.

\subsection{The Trapped Region and the {\em Trapping Boundary}}
The global approach requires 
controlling structures that are not accessible during the evolution.
%, in particular on a given spacetime spatial slice. 
In this context, the 
seminal notion of {\em trapped surface}\cite{Pen65b} plays a crucial role,
capturing the idea that all light rays emitted from the surface locally 
converge. 
Through the singularity theorems and weak cosmic censorship,
it offers a benchmark for the existence of a black hole region:
in strongly predictable spacetimes with proper energy
conditions, trapped surfaces lie inside the black hole region\cite{HawEll73}. 
Moreover, their location does not involve 
%the knowledge of 
a 
whole future  spacetime development.

\subsubsection{Trapped and outer trapped surfaces. Apparent horizons}
\label{s:AH}
Given a closed spatial surface ${\cal S}$ in the spacetime, we can consider the 
light emitted from it along outer and inner directions given, respectively, by null
vectors $\ell^a$ and $k^a$. Then, light locally converges (in the future) 
${\cal S}$ if the area of the emitted light-front spheres decreases in both directions (see though Ref. \refcite{Faraoni:2011zy}).
Denoting the area element ${\cal S}$ as $dA=\sqrt{q}d^2x$, 
the infinitesimal variations of the area along
$\ell^a$ and $k^a$ define outgoing and ingoing expansions $\theta^{(\ell)}$ and 
$\theta^{(k)}$ (see section 
\ref{s:2surfaces} for details)
\be
\label{e:variation_element_area}
\delta_{\ell} \sqrt{q}= \theta^{(\ell)}\sqrt{q} 
\ \ , \ \ 
\delta_{k} \sqrt{q}= \theta^{(k)}\sqrt{q} \ \ .
\ee
A trapped surface is characterized by $ \theta^{(\ell)} \theta^{(k)}>0$.
In the black hole context, in which the singularity occurs in the future,
we refer to ${\cal S}$ as a future {\em trapped surface} (TS) if 
$\theta^{(\ell)}<0,  \theta^{(k)}<0$
and as future {\em marginally trapped surface} (MTS) if one of the expansions,
say $\theta^{(\ell)}$,  vanishes: $\theta^{(\ell)}=0,  \theta^{(k)}\leq 0$.
If a notion of {\em naturally expanding} direction for the light rays
exists (e.g. in isolated systems, the  {\em outer} null direction $\ell^\mu$
pointing to infinity), a related notion of {\em outer trapped  surface}
is given\cite{HawEll73} by  $ \theta^{(\ell)}<0$. 
{\em Marginally outer trapped  surfaces} (MOTS) are characterized by 
$\theta^{(\ell)}=0$.
%% More generally, dropping the global condition on the choice of {\em outgoing}
%% null normal, the requirement of the vanishing of expansion associated with one 
%% of the null normal ${\cal S}$ characterises the {\em marginal surfaces} (MT) 
%% defined by Hayward \cite{Hay04}. We will refer
%% generically to (\ref{e:MOTS}) as the MOTS condition, and the context
%% will make it clear if global conditions are being taken into account
%% or not.

Before proceeding to a characterization of black holes in terms 
of trapped surfaces, 
let us consider trapped surfaces from the perspective of a spatial
slice of spacetime $\Sigma$. 
The {\em trapped region in $\Sigma$}, ${\cal T}_\Sigma\subset\Sigma$,
 is the set of points $p\in \Sigma$ 
belonging to some (outer) trapped surface ${\cal S}\subset\Sigma$. 
The  Apparent Horizon (AH) is then the outermost
boundary of the trapped region ${\cal T}_\Sigma$. A crucial result is 
the following characterization\cite{HawEll73,KriHay97,Chr03}
of AHs: if the trapped 
region ${\cal T}_\Sigma$ in a slice $\Sigma$  
has the structure of a manifold with boundary, the Apparent Horizon is a MOTS, i.e. 
$\theta^{(\ell)}=0$.

Given a 3+1 foliation of spacetime $\{\Sigma_t\}$, let us 
consider the worldtube obtained by piling up the 2-dimensional AHs 
${\cal S}_t\subset \Sigma_t$. Such an AH-worldtube does not need to be a smooth hypersurface 
(it is not even necessarily continuous, as discussed in section \ref{s:jumps}).
This is our first encounter with the notion of a spacetime worldtube foliated by
MOTS. Though these worldtubes are slicing-dependent, their characterization in terms
of MOTSs makes them very useful from a operational perspective.

\subsubsection{The trapped region: definition and caveats}
From a spacetime perspective, no reference to a slice $\Sigma$
must enter into the characterization of the
trapped region. The  spacetime {\em trapped region} ${\cal T}$ is defined 
as the set of points 
$p\in {\cal M}$ belonging to some trapped surface ${\cal S}\subset {\cal M}$.
Its  boundary is referred\cite{Hay94} to as the {\em trapping boundary}.
These concepts offer, in principle, an intrinsically quasi-local avenue 
to address the notion of black hole region and black hole horizon, 
with no reference to asymptotic quantities.

In spite of their appealing features, there are also important
caveats associated with the {\em trapped region} and the {\em trapping boundary}.
In particular, we lack an operational characterization
of the {\em trapping boundary} (see also the contribution by
J.M.M. Senovilla).
A systematic attempt to address this issue is provided by 
the notion of {\em trapping horizon}\cite{Hay94},
namely smooth worldtubes of MOTS (see section \ref{s:trapping_horizons}), 
as a model for the trapping boundary.
Trapping horizons, that are non-unique,
have led to important insights into the structure of the trapped region,
though an operational characterization of the  {\em trapping boundary} is still
missing. 

The difficulties are illustrated in the discussion 
of the relation between the trapping boundary and $\scre$.
In strongly predictable spacetimes
with appropriate energy conditions (see, though Ref.\refcite{Nielsen:2010gm}), 
the trapped region ${\cal T}$ is contained
in the black hole region $\scrb$. In attempts to refine this statement,
support was found\cite{Eard97,SchKri05}
suggesting that the trapping boundary actually coincides with the event horizon, 
though later work\cite{BenDo06} showed that the trapped region not always extend
up to $\scre$. The question is still open for {\em (outer) trapped regions} constructed 
on {\em outer} trapped surfaces, rather than on TSs.
Important insight into these issues 
has been gained in recent works\cite{AmaBenSen10,BenSen10} demonstrating
truly global features of the trapped region ${\cal T}$. In particular:
\bit
\item[i)] The trapping boundary cannot be foliated by MOTS. 
\item[ii)] Closed trapped surfaces can enter into the flat region. This is an important
issue in this approach to black holes, since it was a  main criticism 
in \ref{s:Event_Horizon}.
\item[iii)] Closed trapped surfaces are {\em clairvoyant},
that is, they are {\em aware} of the geometry in non-causally
connected spacetime regions.
This non-local property challenges their applicability 
for an operational characterization of black holes.
\eit

\subsection{A {\em pragmatic approach} to quasi-local black hole horizons}
Trapping horizons offer a sound avenue towards the  quasi-local 
understanding of black hole physics. 
They provide crucial insight in gravitational scenarios where a quasi-local notion of
black hole horizon is essential, such as black hole
thermodynamics beyond equilibrium, the characterization of physical parameters 
of strongly dynamical astrophysical black holes (notably
in numerical simulations), semi-classical collapse, quantum
gravity or mathematical relativity (cf. A. Nielsen's contribution). 
But, on the other hand, 
issues like their non-uniqueness or the {\em clairvoyant} properties of 
trapped surfaces pose fundamental questions that cannot be ignored.

We do not aim here at addressing first-principles questions about the
role of trapping horizons as a characterization of black hole horizons.
We rather assume a {\em pragmatic approach} to the study
of gravitational dynamics, which underlines the role of trapping horizons
as hypersurfaces of remarkable geometric properties 
in black hole spacetimes. More specifically, our main interests are:
\bit
\item[i)] The construction and  diagnosis of
black hole spacetimes in Initial (Boundary) Value Problem approach.
\item[ii)] Identification of a geometric probe into near-horizon spacetime dynamics.
\eit
Point ii) is particularly important in the 
study of gravity in the strong-field regime, where the lack of {\em rigid structures}
(e.g. symmetries, a {\em background} spacetime...) is a generic and
essential problem. 
Given our interests and the adopted pragmatic methodology, we look for a geometric
object such that: a) represents a footprint
of black holes, providing a probe into their
geometry; b) is adapted, by construction, to an Initial-Boundary Value 
Problem approach; and c) although not-necessarily unique, provides
 a geometric structure with some sort of {\em rigidity} property.
As we shall see in the following, dynamical trapping horizons fulfill these requirements.

\subsection{General scheme}
In section \ref{s:concepts_definitions} we introduce the basics of the
geometry of closed surfaces in a Lorentzian manifold and motivate
quasi-local horizons in stationary and dynamical regimes.
Section \ref{s:Properties} reviews their geometric
properties and their special features  as 
{\em physical} boundaries. Sections \ref{s:BH_IVP} and \ref{s:BH_aposteriori} are 
devoted to applications in a 3+1 description of the spacetime. 
Section  \ref{s:BH_IVP} shows the use of quasi-local horizons as inner boundary conditions
for elliptic equations in General Relativity,
whereas section \ref{s:BH_aposteriori} discusses some applications to the analysis 
of spacetimes, in particular their role in a {\em correlation approach} to
spacetime dynamics. In section \ref{s:Gen_pers} a general overview is presented.

%%%%%%%%%%%%%%%%%%%%%%%%%%%%%%%%%%%%%%%%%%%%%%%%%%%%%%%%%%%%%%%%%%%%%%%%%%
%%%%%%%%%%%%%%%%%%%%%%%%%%%%%%%%%%%%%%%%%%%%%%%%%%%%%%%%%%%%%%%%%%%%%%%%%%
%%%%%%%%%%%%%%%%%%%%%%%%%%%%%%%%%%%%%%%%%%%%%%%%%%%%%%%%%%%%%%%%%%%%%%%%%%
\section{Quasi-local horizons: Concepts and Definitions}
\label{s:concepts_definitions}
%%%%%%%%%%%%%%%%%%%%%%%%%%%%%%%%%%%%%%%%%%%%%%%%%%%%%%%%%%%%%%%%%%%%%%%%%%
%%%%%%%%%%%%%%%%%%%%%%%%%%%%%%%%%%%%%%%%%%%%%%%%%%%%%%%%%%%%%%%%%%%%%%%%%%
%%%%%%%%%%%%%%%%%%%%%%%%%%%%%%%%%%%%%%%%%%%%%%%%%%%%%%%%%%%%%%%%%%%%%%%%%%

\subsection{Geometry of spacelike closed 2-surfaces ${\cal S}$}
\label{s:2surfaces}
\subsubsection{Normal plane: {\em outgoing} and {\em ingoing} null
vectors}
Let us consider a spacetime $({\cal M}, g_{ab})$ with 
Levi-Civita connection $\nabla_a$.
Given a spacelike closed (compact without boundary) 
2-surface ${\cal S}$ in ${\cal M}$
and a point $p\in {\cal S}$, the tangent space
splits as $T_p{\cal M} = T_p{\cal S} \oplus T_p^\perp {\cal S}$.
We span the normal plane  $T_p^\perp {\cal S}$ either by (future-oriented)
null vectors $\ell^a$ and $k^a$ (defined by the intersection between $T_p^\perp {\cal S}$
and the null cone at $p$) or by any pair of normal timelike vector $n^a$ 
and spacelike vector $s^a$. Let us denote conventionally $\ell^a$ to be the {\em outgoing}
null normal and $k^a$ the {\em ingoing} one. We choose normalizations:
\bea
\label{e:normalization}
\ell^a \ell_a = 0  \ , \  k^a k_a = 0  \ , \ \ell^a k_a = -1 \ \  , \ \
n^a n_a = -1 \ , \  s^a s_a = 1  \ , \ n^a s_a = 0  \ ,
\eea
Directions $\ell^a$ and $k^a$ are uniquely determined,
but a {\em normalization-boost} freedom 
\bea
\label{e:boost_freedom}
\ell'^a = f \ell^a  \ &,& \ k'^a = f^{-1} k^a   \\
n'^a = \mathrm{cosh}(\sigma) n^a + \mathrm{sinh}(\sigma) s^a  \ &,& \ 
s'^a = \mathrm{sinh}(\sigma) n^a + \mathrm{cosh}(\sigma) s^a  \nn \ ,
\eea
remains for some arbitrary rescaling positive function $f$ on ${\cal S}$ 
(where  $\sigma = \mathrm{ln}(f)$ and $\ell^a = \lambda(n^a + s^a)/\sqrt{2}$ and 
$k^a = \lambda^{-1}(n^a - s^a)/\sqrt{2}$, for some function $\lambda$ on ${\cal S}$).

\subsubsection{Intrinsic geometry of ${\cal S}$ }
The induced metric on ${\cal S}$ is given by
\be
\label{e:metric_S}
q_{ab}=g_{ab}+k_a \ell_b + \ell_a k_b = g_{ab} +n_a n_b - s_a s_b \ ,
\ee
so that ${q^a}_b$ is the projector onto ${\cal S}$
\be
\label{e:projector_S}
{q^a}_b {q^b}_c = {q^a}_c \ \ , \ \ {q^a}_b v^b = v^a (\forall v^a\in T{\cal S})
\ \ , \ \ {q^a}_b w^b = 0 (\forall w^a\in T^\perp {\cal S}) \ .
\ee
We denote the Levi-Civita connection associated with $q_{ab}$ as
${}^2\!D_a$. The volume form on ${\cal S}$ will be denoted
by ${}^2\!\epsilon = \sqrt{q} dx^1\wedge dx^2$, i.e.
${}^2\!\epsilon_{ab}= n^c s^d {}^4\!\epsilon_{cdab}$, though we will 
also employ the area measure notation $dA=\sqrt{q}d^2x$.

\subsubsection{Extrinsic  geometry of ${\cal S}$ in $({\cal M},g)$}
\label{s:extrinsic_S}
We define the {\em second fundamental tensor} of $({\cal S}, q_{ab})$ in 
$({\cal M}, g_{ab})$ (also, {\em shape tensor} or {\em extrinsic curvature tensor})
as
\be
\label{e:second_fund_form_S}
{\cal K}^c_{ab}= {q^d}_a {q^e}_b \nabla_d {q^c}_e \ \ ,
\ee
where $c$ is an index in the normal plane $T^\perp{\cal S}$,
whereas $a$ and $b$ are indices in $T{\cal S}$.
Given a vector $v^a$ normal to ${\cal S}$, we can define the 
{\em deformation tensor} $\Theta^{(v)}_{ab}$ as
\be
\label{e:deformation_tensor}
\Theta^{(v)}_{ab}=  {q^c}_a {q^d}_b \nabla_c v_d \ \ .
\ee
Then, using expression (\ref{e:metric_S}), the {\em second fundamental tensor} 
can be expressed as
\bea
\label{e:2nd_fund_form}
 {\cal K}^c_{ab} &=& k^c \Theta^{(\ell)}_{ab} + \ell^c \Theta^{(k)}_{ab} 
= n^c \Theta^{(n)}_{ab} -  s^c \Theta^{(s)}_{ab} \ .
\eea
We can express $\Theta^{(v)}_{ab}$ in terms of the variation of the intrinsic metric
along $v^a$.
Given a (tensorial) object ${A_{a_1...a_n}}^{b_1...b_m}$ tangent to ${\cal S}$
we denote by $\delta_v$ the operator ${(\delta_v A)_{a_1...a_n}}^{b_1...b_m}=
{q_{a_1}}^{c_1}...{q_{a_n}}^{c_n}
{q_{d_1}}^{b_1}...{q_{d_m}}^{b_m}{\cal L}_v {A_{c_1...c_n}}^{d_1...d_m}$, where 
${\cal L}_v$ denote the Lie derivative along (some extension of) $v^a$. Then, it follows
\bea
\label{e:delta_q}
\delta_v q_{ab} = \frac{1}{2}\Theta^{(v)}_{ab} \ .
\eea

\noi {\em a) Shear and expansion associated with $v^a$}. Defining
the expansion $\theta^{(v)}$ and shear tensor $\sigma^{(v)}_{ab}$ associated 
with the normal vector $v^a$ as
\be
\label{e:expansion_shear}
\theta^{(v)} \equiv q^{ab}\nabla_av_b = \delta_v \mathrm{ln}\sqrt{q} \ \ , \ \ 
\sigma^{(v)}_{ab} \equiv \Theta^{(v)}_{ab} - \frac{1}{2} \theta^{(v)} q_{ab} \ \ ,
\ee
we express the deformation tensor $\Theta^{(v)}_{ab}$ in terms of
his trace and traceless parts
\bea
\label{e:trace_traceless_Theta}
\Theta^{(v)}_{ab} = \sigma^{(v)}_{ab} +  \frac{1}{2} \theta^{(v)} q_{ab} \ .
\eea
\noi {\em b) Mean curvature vector $H^a$}. Taking the trace of $\Theta^{(v)}_{ab}$
on ${\cal S}$ we define the 
{\em mean curvature vector}\footnote{\label{f:sign_convention_shape}Note the opposite 
sign convention with respect to the contribution by J.M.M. Senovilla.}  
\be
\label{e:mean_curvature_vector}
H^c \equiv  q^{ab}{\cal K}^c_{ab} = \theta^{(\ell)} k^c 
+  \theta^{(k)} \ell^c \ \ .
\ee
The extrinsic curvature information of $({\cal S}, q_{ab})$ 
in $({\cal M}, g_{ab})$ is completed by the {\em normal fundamental forms}
associated with normal vectors $v^a$. In particular\cite{Gou05}
\bea
\Omega^{(n)}_a = s^c {q^d}_a \nabla_d n_c \ \ &,& \ \ \Omega^{(s)}_a = n^c {q^d}_a  \nabla_d s_c  \nn\\
\Omega^{(\ell)}_a = \frac{1}{k^b \ell_b}k^c {q^d}_a \nabla_d \ell_c \ \ &,& \ \ 
\Omega^{(k)}_a = \frac{1}{k^b \ell_b}\ell^c {q^d}_a   \nabla_d k_c \ .
\eea
All these normal fundamental forms are related up to a sign and a total derivative
on ${\cal S}$. Using the
normalizations (\ref{e:normalization}) we get\footnote{\label{f:delta_Omega}When
using $\ell^ak_a=-e^\sigma$ one gets: $ \Omega^{(\ell)}_a 
= -  \Omega^{(k)}_a -{}^2\!D_a\sigma $. This will be relevant later, 
in Eq. (\ref{e:delta_tildek}).}: $\Omega^{(n)}_a = -  \Omega^{(s)}_a, 
\Omega^{(\ell)}_a = -  \Omega^{(k)}_a,
 \Omega^{(\ell)}_a = \Omega^{(n)}_a - {}^2\!D_a\lambda$.
We choose to employ the 1-form $\Omega^{(\ell)}_a$ in the following.

\subsubsection{Transformation properties under null normal rescaling}
Under the rescaling (\ref{e:normalization}) $\ell^a \to f \ell^a$, $k^a \to f^{-1} k^a$
the introduced fields transform as 
\bea
\label{e:rescaling}
\begin{array}{rclcrclcrcl}
q_{ab}  &\to& q_{ab} & \qquad & {}^2\!D_a&\to& {}^2\!D_a  & & \qquad & &\\
{\cal K}^c_{ab}&\to&{\cal K}^c_{ab}& \qquad & H^a &\to& H^a  && \qquad && \\
 \Theta^{(\ell)}_{ab} &\to& f \Theta^{(\ell)}_{ab} & \qquad&
\theta^{(\ell)} &\to& f \theta^{(\ell)} & & \qquad
 \sigma^{(\ell)}_{ab} &\to& f \sigma^{(\ell)}_{ab}   \\
 \Theta^{(k)}_{ab} &\to& f^{-1} \Theta^{(k)}_{ab} & \qquad&
\theta^{(k)} &\to& f^{-1} \theta^{(k)} & & \qquad
 \sigma^{(k)}_{ab} &\to& f^{-1} \sigma^{(k)}_{ab}   \\
\Omega^{(\ell)}_a  &\to& \Omega^{(\ell)}_a + {}^2\!D_a(\mathrm{ln}f) & \qquad\qquad &&&&
& \qquad&&
\end{array}
\eea
Finally, given an axial Killing vector $\phi^a$ on ${\cal S}$, we
can write the angular momentum\footnote{The quantity $J$ coincides with the Komar
angular momentum in case that $\phi^a$ can be extended to an axial Killing in the 
neighbourhood of ${\cal S}$.}
\bea
\label{e:Komar_angular_momentum}
J = \frac{1}{8\pi} \int_{\cal S} \Omega^{(\ell)}_a \phi^a {}^2\!\epsilon \ .
\eea
The transformation rule of $\Omega^{(\ell)}_a$ in (\ref{e:rescaling})
together with the divergence-free property of $\phi^a$ (following from
its Killing character) guarantee that the quantity $J$ does not depend on the choice of 
null normals $\ell^a, k^a$ (i.e. $J$ does not change under a boost).

\subsection{Trapping Horizons}
\label{s:trapping_horizons}

\subsubsection{Worldtubes of marginally trapped surfaces}
A {\em trapping horizon}\cite{Hay94}
is (the closure of) a hypersurface ${\cal H}$ 
foliated by closed marginal (outer) trapped surfaces:
$\Hor = \bigcup_{t\in\mathbb{R}} \Sp_t $,  with 
$\left.\theta^{(\ell)}\right|_{{\cal S}_t}=0$. 
Trapping horizons
are classified according to the signs of $\theta^{(k)}$
and $\delta_{k}\theta^{(\ell)}$. In particular, the sign of $\theta^{(k)}$ 
controls if the singularity occurs either in 
 the  {\em future} or in the {\em past} of ${\cal S}$, whereas
the sign of $\delta_{k}\theta^{(\ell)}$ controls 
the (local) {\em outer-} or {\em innermost} character of ${\cal H}$.
Then, a trapping horizon is said to be: i) {\em future} 
(respectively, {\em past}) if $\theta^{(k)}<0$ (respectively, $\theta^{(k)}>0$),
and ii) {\em outer} (respectively, {\em inner}) if there exists\footnote{The sign 
of $\delta_{k}\theta^{(\ell)}$ is not invariant on the whole
${\cal S}$ under a rescaling (\ref{e:normalization}). However, if 
there exists $\ell^a$ and $k^a$
such that  $\delta_{k}\theta^{(\ell)}<0$ on ${\cal S}$, then there does not exist any choice of
$\ell^a$ and $k^a$ such that $\delta_{k}\theta^{(\ell)}>0$ on ${\cal S}$; see
Ref. \refcite{BooFai07} and also the marginally trapped surface stability condition 
in Ref. \refcite{Racz:2008tf}.}
 $\ell^a$ and $k^a$
such that  $\delta_{k}\theta^{(\ell)}<0$ (respectively, $\delta_{k}\theta^{(\ell)}>0$).

\subsubsection{Future Outer Trapping Horizons}
In a black hole setting the singularity occurs in the future of sections ${\cal S}_t$
of ${\cal H}$, so that the
related trapping horizon is of {\em future} type, $\theta^{(k)}<0$.
In addition, when considering displacements along $k^a$ ({\em ingoing} direction)
we should move 
into the trapped region, i.e. $\delta_{k}\theta^{(\ell)}<0$, so that the trapping horizon
should be {\em outer}. 

The resulting characterization of quasi-local black hole  
horizons as {\em Future Outer Trapping Horizons} (FOTHs) is further supported by 
the following 
analysis of the area evolution. Hawking's area theorem for event horizons 
(cf. section \ref{s:Event_Horizon}) captures a fundamental feature
of classical black holes. It is natural to wonder about a quasi-local version
of it.
Let us consider an evolution vector $h^a$ along the trapping horizon 
${\cal H}$, characterized as: i)
$h^a$ is tangent to $\Hor$ and orthogonal to ${\cal S}_t$, and ii)
$h^a$ transports ${\cal S}_t$ onto ${\cal S}_{t+\delta t}$: ${\delta}_h t =1$. 
We can write $h^a$ and a {\em dual} vector $\tau^a$ orthogonal to ${\cal H}$ as
\bea
\label{e:h_m}
h^a = \ell^a - C k^a \ \ , \ \ \tau^a = \ell^a + C k^a \ .
\eea
Then $h^a h_a = -\tau^a \tau_a = 2C$, i.e. $h^a$ is 
spacelike for $C>0$, null for $C=0$ and timelike for $C<0$.
The evolution of the area $A= \int_{\cal S} dA= \int_{\cal S} {}^2\!\epsilon$
along $h^a$ is given by
\bea
\label{e:area_change}
\delta_{h} A = \int_{\cal S} \theta^{(h)}{}^2\!\epsilon
= \int_{\cal S} \left(\theta^{(\ell)} - C \theta^{(k)}\right){}^2\!\epsilon =
- \int_{\cal S} C \theta^{(k)} {}^2\!\epsilon \ .
\eea
Considering for simplicity the spherical symmetric case
($C=\mathrm{const}$; see discussion of Eq. (\ref{e:TH_condition}) 
in \ref{s:DH_eneric_properties}, for the general case), 
the {\em trapping horizon} condition, $\delta_h \theta^{(\ell)} = 0$, writes
$\delta_\ell \theta^{(\ell)} - C \delta_k \theta^{(\ell)}=0$, so that 
$C = \frac{\delta_\ell \theta^{(\ell)}}{\delta_k \theta^{(\ell)}}$.
Applying the {\em Raychaudhuri} equation for $\delta_\ell \theta^{(\ell)}$
[see later Eq. (\ref{e:null_Raychaudhuri})], 
together with the $\theta^{(\ell)}=0$ condition, we find
\be
\label{e:C_spherical}
C = - \frac{\sigma_{ab}^{(\ell)}{\sigma^{(\ell)}}^{ab}+ 8\pi T_{ab}\ell^a\ell^b}
{\delta_k \theta^{(\ell)} } \ \ .
\ee
Under the null energy and outer horizon conditions, it follows
$C\geq 0$, so that the future condition guarantees the non-decrease of the area
in (\ref{e:area_change}).
Therefore,  FOTHs
are {\em null} or {\em spacelike} hypersurfaces ($C\geq 0$), satisfying
an area law result, and therefore providing
appropriate models for quasi-local black hole horizons. 

\subsection{Isolated and Dynamical Horizons}
The  distinct geometric structure of
null and spatial hypersurfaces suggests different 
strategies for the study of the stationary and dynamical regimes of quasi-local 
black holes, modeled as future outer trapping horizons. This has led to the 
parallel development of 
{\em isolated  horizon} and the {\em dynamical horizon} 
frameworks\cite{AshKri03,Boo05,GouJar06,Krishnan:2007va}.

In equilibrium, {\em Isolated Horizons} (IH) provide a hierarchy of 
geometric structures constructed on a null hypersurface ${\cal H}$
that is foliated by closed (outer) marginally trapped surfaces. They characterize
different levels of stationarity for a black hole horizon 
in an otherwise dynamical environment: \\
\noi i) {\em Non-Expanding Horizons} (NEH). They represent the minimal 
notion of equilibrium by imposing the stationarity of the intrinsic geometry $q_{ab}$.\\
\noi ii) {\em Weakly Isolated Horizons} (WIH). They are NEHs endowed with an 
additional structure needed for a 
Hamiltonian analysis of the horizon and its related (thermo-)dynamics.
They impose no additional constraints on the geometry of the NEH. \\
\noi iii) {\em Isolated Horizons} (IH). These are WIHs whose extrinsic
geometry is also invariant along the evolution. They provide the strongest 
 stationarity notion on ${\cal H}$.

The non-stationary regime can be characterized by {\em Dynamical Horizons} (DH),
namely spacelike hypersurfaces ${\cal H}$  foliated by closed future marginally
trapped surfaces, i.e. $ \theta^{(\ell)}=0$ and $ \theta^{(k)}<0$.
Introduced in a  3+1 formulation, they provide a complementary perspective 
to the {\em dual-null foliation formulation}\cite{Hay94} of trapping horizons,
making them naturally adapted for
an Initial Value Problem perspective.

\subsection{IHs and DHs as stationary and dynamical sections of FOTHs}
\label{s:equivalence}
A natural question when considering the transition from equilibrium to the dynamical 
regime is whether a section ${\cal S}_t$ of a FOTH can be partially stationary
and partially dynamical. Or, in other words, whether the element of area $dA$
can be non-expanding ($C=0$) in a part of  ${\cal S}_t$ whereas it already expands 
($C>0$) in another part. Namely, can $h^a$ 
be both null and spacelike on a section  ${\cal S}_t$ of a FOTH?

The answer is in the negative.
Transitions between non-expanding and dynamical
parts of a FOTH must happen {\em all at once}.
More precisely, assuming the null energy condition, a FOTH
can be completely partitioned into non-expanding and dynamical sections.
For a section ${\cal S}_t$ to be completely dynamical ($C>0$) it suffices that it has
$\delta_\ell \theta^{(\ell)}<0$ somewhere on it. Otherwise $h^a$ is null
($C=0$) all over ${\cal S}_t$\cite{AndMarSim05,BooFai07}.

In more physical terms, it suffices that some {\em energy} 
crosses the horizon {\em somewhere}, and the {\em whole} horizon instantaneously
grows as a whole. 
This non-local behaviour is a consequence of the {\em elliptic}
nature of quasi-local horizons. As shown in section \ref{e:FOTH_charac_prop},
the function $C$ determining the metric type of $h^a$ satisfies
an elliptic equation [cf. Eq. (\ref{e:TH_condition})].
Under the  outer condition $\delta_k\theta^{(\ell)}<0$
one can apply
a {\em maximum principle} to show that  $C$ is non-negative [generalization
of Eq. (\ref{e:C_spherical})]. Moreover, it suffices
that $ \delta_\ell\theta^{(\ell)}\neq 0$ somewhere, for having $C>0$ 
everywhere.

%%%%%%%%%%%%%%%%%%%%%%%%%%%%%%%%%%%%%%%%%%%%%%%%%%%%%%%%%%%%%%%%%%%%%%%%%%
%%%%%%%%%%%%%%%%%%%%%%%%%%%%%%%%%%%%%%%%%%%%%%%%%%%%%%%%%%%%%%%%%%%%%%%%%%
%%%%%%%%%%%%%%%%%%%%%%%%%%%%%%%%%%%%%%%%%%%%%%%%%%%%%%%%%%%%%%%%%%%%%%%%%%
\section{Quasi-local horizons: properties from a  3+1 perspective}
\label{s:Properties}
%%%%%%%%%%%%%%%%%%%%%%%%%%%%%%%%%%%%%%%%%%%%%%%%%%%%%%%%%%%%%%%%%%%%%%%%%%
%%%%%%%%%%%%%%%%%%%%%%%%%%%%%%%%%%%%%%%%%%%%%%%%%%%%%%%%%%%%%%%%%%%%%%%%%%
%%%%%%%%%%%%%%%%%%%%%%%%%%%%%%%%%%%%%%%%%%%%%%%%%%%%%%%%%%%%%%%%%%%%%%%%%%

\subsection{Equilibrium regime}
\label{s:equilibrium_regime}

\subsubsection{Null hypersurfaces: characterization and basic elements}
A hypersurface ${\cal H}$ is null if and only if the induced metric
is degenerated. Equivalently, if and only if there is 
a tangent null
vector $\ell^a$ orthogonal to all vectors tangent to ${\cal H}$: 
$\ell^a v_a=0, \; \forall v^a \in T{\cal H}$.

Let us introduce some elements on the geometry of ${\cal H}$.
Choosing a null vector $k^a$ transverse to ${\cal H}$,
we can write\footnote{We abuse notation and employ the same notation
employed in sections ${\cal S}_t$ of ${\cal H}$, cf. Eq. (\ref{e:metric_S}).} 
the degenerated metric as $q_{ab}= g_{ab}+k_a\ell_b + \ell_a k_b$. 
A projector onto ${\cal H}$ can also 
be constructed as: ${\Pi_a}^b = {\delta_a}^b + \ell_ak^b=
{q_a}^b - k_a\ell^b$. 
As a part of the extrinsic curvature of ${\cal H}$, a 
{\em  rotation 1-form} can be introduced\cite{AshBeeLew01} on ${\cal H}$ as
 $\omega^{(\ell)}_a = \frac{1}{\ell^a k_a}k^c \nabla_a \ell_c$. This 1-form 
{\em lives} on ${\cal H}$, i.e. $k^a \omega^{(\ell)}_a = 0$. In particular, we can write 
${\Pi_a}^c\nabla_c \ell^b = \omega^{(\ell)}_a \ell^b + {{\Theta^{(\ell)}}_a}^b$,
where  $\Theta^{(\ell)}_{ab}$ is given by expression (\ref{e:deformation_tensor})
[cf. Eq. (5.23) in Ref. \refcite{GouJar06}]. Contracting with $\ell^a$ we find: 
$\ell^c \nabla_c \ell^a = \kappa^{(\ell)} \ell^a$, a pre-geodesic equation where
the non-affinity coefficient $\kappa^{(\ell)}$ is defined as
$\kappa^{(\ell)} = \ell^a \omega^{(\ell)}_a$.
If a foliation $\{{\cal S}_t\}$ of ${\cal H}$ is given, we can write
[cf. Eq. (5.35) in Ref. \refcite{GouJar06}]:  
$\omega^{(\ell)}_a= \Omega^{(\ell)}_a -  \kappa^{(\ell)} k_a$.

Vectors $\ell^a$ and $k^a$ can be completed
to a tetrad $\{ \ell^a, k^a, (e_1)^a, (e_2)^a \}$, where $(e_i)^a$ are tangent
to sections ${\cal S}_t$. Normalizations given in (\ref{e:normalization}) are 
then completed to
\be
\ell \cdot {(e_i)}_{a} = 0 \ , \  k^a {(e_i)}_{a} = 0 \ , {(e_i)}^a {(e_i)}_{b} = 
\delta_{ab} \ .
\ee
Defining the complex null vector
$m^a = \frac{1}{\sqrt{2}}[(e_1)^a+ i (e_2)^a]$, the Weyl scalars are defined
as the components of the Weyl tensor $C^a_{\ \, bcd}$ in the null tetrad  
$\{ \ell^a, k^a, m^a, \overline{m}^a \}$
\be
\label{e:Weyl_scalars}
	\begin{array}{ll}
\Psi_0 = C^a_{\ \, bcd}\; \ell_a m^b \ell^c m^d & 
\qquad \Psi_3 = C^a_{\ \, bcd}\; \ell_a k^b \overline{m}^c k^d \\
\Psi_1 = C^a_{\ \, bcd}\; \ell_a m^b \ell^c k^d &
\qquad \Psi_4 = C^a_{\ \, bcd}\; \overline{m}_a k^b \overline{m}^c k^d \\
\Psi_2 = C^a_{\ \, bcd}\; \ell_a m^b \overline{m}^c k^d &
    	\end{array}
\ee

\subsubsection{Null hypersurfaces: evolution}
It is illustrative to give a 3+1 perspective on ${\cal H}$. 
Given a foliation $\Hor = \bigcup_{t\in\mathbb{R}} \Sp_t $
let us evaluate explicitly the evolution along $\ell^a$ of quantities 
defined on sections $\Sp_t$. 

\noi i) Expansion equation (null Raychaudhuri equation):
\be
\label{e:null_Raychaudhuri}
\delta_\ell \theta^{(\ell)} - \kappa^{(\ell)}\theta^{(\ell)} 
	+  {1\over 2} {\theta^{(\ell)}}^2 + \sigma^{(\ell)}_{ab} {\sigma^{(\ell)}}^{ab}  
	+ 8\pi T_{ab}\ell^a\ell^b = 0 \ .
\ee

\noi ii) Tidal equation:
\be
\label{e:null_tidal}
\delta_\ell\sigma^{(\ell)}_{ab}
=\kappa^{(\ell)} \, \sigma^{(\ell)}_{ab}
+ \sigma^{(\ell)}_{cd}{\sigma^{(\ell)}}^{cd} \, q_{ab}
 -q^c_{\ \, a} q^d_{\ \, b}    C_{ecfd} \ell^e\ell^f  \ .
\ee

\noi iii) Evolution for $\Omega_a$:
\be
\label{e:null_Omega}
\delta_\ell \Omega^{(\ell)}_c + \theta^{(\ell)} \, \Omega^{(\ell)}_a =
       8\pi T_{cd} \, \ell^c q^d_{\ \, a}
       +  {}^2\!D_a \left( \kappa^{(\ell)} + \frac{\theta^{(\ell)}}{2} \right)
       -  {}^2\!D_c {\sigma^{(\ell)}}^c_{\ \, a}  \ .
\ee

\subsubsection{Non-Expanding Horizons}
\label{s:NEH}
A NEH \cite{AshKri04} is
a null-hypersurface ${\cal H}\approx S^2 \times \mathbb{R}$, on which 
the expansion associated with $\ell^a$ vanishes
($\theta^{(\ell)} = 0$), the 
Einstein equations hold and 
$-{T^a}_c\ell^c$ is future directed ({\em null dominant energy condition}).
Note that any foliation $\Hor = \bigcup_{t\in\mathbb{R}} \Sp_t $
produces a foliation of ${\cal H}$  by MOTS  $\Sp_t $.

\medskip
\noi i) {\em  NEH characterization}.
Making $\theta^{(\ell)}=0$ in the Raychaudhuri Eq. (\ref{e:null_Raychaudhuri})
we get
\be
 \sigma_{ab} \sigma^{ab} + 8\pi T_{ab}\ell^a\ell^b = 0 \ .
\ee
Since the two terms are positive-definite,  they vanish independently.
This provides an {\em instantaneous} characterization of a NEH:
\be
\label{e:NEH_condition}
\theta^{(\ell)}=0 \ \ , \ \ \sigma^{(\ell)}_{ab}=0 \ \ , \ \ T_{ab}\ell^a\ell^b=0 \ .
%\qquad
% {\cal K}^c_{ab} = k^c \Theta^{(\ell)}_{ab} + \ell^c \Theta^{(k)}_{ab}  \nn
\ee
From Eq. (\ref{e:trace_traceless_Theta}) with $v^a=\ell^a$, 
it follows $\Theta^{(\ell)}_{ab} = 0$. The NEH
characterization is equivalent, cf. Eq. (\ref{e:delta_q}), 
to the {\em evolution independence}
 of the induced metric $q_{ab}$
\bea
\label{e:Theta_zero}
\delta_\ell q_{cd} = \frac{1}{2}\Theta^{(\ell)}_{ab} = 0 \ .
\eea
From Eq. (\ref{e:2nd_fund_form}), we conclude that a NEH 
fixes half of the degrees of freedom in the second fundamental form ${\cal K}^c_{ab}$
of ${\cal S}_t$ in ${\cal M}$. This will be relevant in section \ref{s:ID_NEH}.

\medskip
\noi ii) {\em Connection $\hat{\nabla}_a$ on a NEH}. 
A null hypersurface has no unique (Levi-Civita) connection 
compatible with the metric. However, on a NEH ${\cal H}$
one can introduce a preferred connection as that one
induced from the spacetime connection $\nabla_a$: 
$u^c\hat{\nabla}_cw^a=u^c\nabla_cw^a$, $\forall u^a, w^a\in T{\cal H}$. 
Indeed using NEH characterization 
(\ref{e:Theta_zero}),  $u^c\nabla_cw^a$ is tangent to ${\cal H}$:
$\ell_d (u^c\nabla_cw^d) =u^c\nabla_c(\ell_dw^d)-
 u^cw^d \Theta^{(\ell)}_{cd}=0$.

\medskip
\noi iii) {\em Geometry of a NEH}. 
We refer\cite{AshBeeLew02} to the pair $(q_{ab}, \hat{\nabla}_a)$
as the {\em geometry of a NEH}.
Writing the components of the $\hat{\nabla}_a$ connection
in terms of quantities on ${\cal S}_t$
\bea
q^c_{\ \, a} \, q^b_{\ \, d} \hat{\nabla}_c v^d &=& 
    {}^2\!D_a (q^b_{\ \, c} v^c) q^c_{\ \, a} \nn\\
k_d  \hat{\nabla}_c v^d &=& 
    {}^2\!D_a (v^c k_c) 
 - q^c_{\ \, a} v^d \Theta^{(k)}_{cd}  \\
 \ell^c  \hat{\nabla}_c v^a &=& \delta_{\ell}v^a + v^c \omega^{(\ell)}_c
 \ell^a \nn \ ,
\eea
 the free data on a NEH are given, from an evolution perspective, by
$(q_{ab}|_{{\mathcal S}_t}, 
\Omega^{(\ell)}_a|_{{\mathcal S}_t}, \kappa^{(\ell)}|_{\Hor}, 
\Theta^{(k)}_{ab}|_{{\mathcal S}_t})$, 
where $q_{ab}$ is time independent. 

\medskip
\noi iv){\em Weyl tensor on a NEH}. 
Under the rescaling (\ref{e:boost_freedom}),
the 1-form $\omega^{(\ell)}_a$ transforms as
$\omega^{(\ell)}_a \to \omega^{(\ell)}_a + 
\hat{\nabla}_a\mathrm{ln}f$. Its exterior derivative $d\omega^{(\ell)}$ provides 
a gauge invariant object: understanding $\omega^{(\ell)}_a$
as a gauge connection, $d\omega^{(\ell)}$ is its gauge-invariant curvature.
Using the NEH condition, $\Theta^{(\ell)}_{ab}=0$, one can 
express (cf. section 7.6.2. in Ref. \refcite{GouJar06})
\be
d\omega^{(\ell)} = 2\; \mathrm{Im}\Psi_2 {}^2\!\epsilon \ .
\ee
Hence, $\mathrm{Im}\Psi_2$ is gauge invariant on a NEH. Actually
the full $\Psi_2$ is invariant, as it follows from its
{\em boost} transformation rules and the values of $\Psi_0$ and $\Psi_1$
on a NEH\cite{GouJar06},
\be
\label{e:Psi0_Psi1}
\Psi_0|_{\Hor}=\Psi_1|_{\Hor}=0 \ . 
\ee

\subsubsection{Weakly Isolated Horizons}
\label{s:WIH}
A {\em Weakly Isolated Horizon} (WIH) $({\cal H}, [\ell^a])$
is a NEH together with a class of null normals $[\ell^a]$
such that: $\delta_{\ell}\omega^{(\ell)}_a=0$. This condition permits to set a well-posed
variational problem for spacetimes containing stationary quasi-local
horizons. This enables the development of a Hamiltonian analysis on the horizon
${\cal H}$
leading to the construction of conserved quantities
under WIH-symmetries\cite{AshBeeLew01}.  In particular, the expression for
the angular momentum in Eq. (\ref{e:Komar_angular_momentum})  is recovered
\bea
J_{\mathcal H} = 
\frac{1}{8\pi}\int_{\Sp_t} \omega^{(\ell)}_c \phi^c  \; {}^2\epsilon
=\frac{1}{8\pi}\int_{\Sp_t} \Omega^{(\ell)}_c \phi^c  \; {}^2\epsilon 
=-\frac{1}{4\pi}\int_{\Sp_t} f \mathrm{Im}\Psi_2 {}^2\epsilon  \ ,
\eea
with $\phi^a= {}^2\!D_cf \;{}^2\!\epsilon^{ac}$ ($\phi^a$ is an axial
Killing vector, in particular divergence-free).

The WIH structure is relevant for the discussion of IH thermodynamics
 (cf. A. Nielsen's contribution).
We do not address this issue here  and just comment
on the equivalence of the WIH condition with a thermodynamical {\em zeroth law}.
Reminding $\omega^{(\ell)}_a=\Omega^{(\ell)}_a - \kappa^{(\ell)} k_a$,
the (vacuum) evolution equation (\ref{e:null_Omega}) for $\Omega^{(\ell)}_a$ leads to
${\cal L}_{\ell}\Omega^{(\ell)}_a = {}^2\!D_a\kappa^{(\ell)}$. More generally,
${\delta}_{\ell}\omega^{(\ell)}_a = \hat{\nabla}\kappa^{(\ell)}$ 
(cf. for example Eq. (8.5) in Ref. 
\refcite{GouJar06}). That is, on WIHs
the non-affinity coefficient ({\em surface gravity}) is constant: $\kappa^{(\ell)}=\kappa_o$.

\medskip
\noi {\em WIHs and NEH geometry}. 
 WIHs do not constraint the underlying NEH geometry. In other
words, every NEH admits a WIH structure. In fact, given
$\kappa^{(\ell)}\neq \mathrm{const}$, the rescaling $\ell'= \alpha \ell$,
with $ \kappa_o=\mathrm{const}= \nabla_\ell \alpha + \alpha \kappa^{(\ell)}$,
leads to a constant $\kappa^{(\ell')}=\kappa_o$.
Finally, free data for a WIH are again $(q_{ab}|_{{\mathcal S}_t}, 
\Omega^{(\ell)}_a|_{{\mathcal S}_t}, \kappa^{(\ell)}|_{\Hor}, 
\Theta^{(k)}_{ab}|_{{\mathcal S}_t})$,  but now $q_{ab}|_{{\mathcal S}_t}$,
$\Omega^{(\ell)}_a|_{{\mathcal S}_t}$ and $\kappa^{(\ell)}|_{\Hor}=\kappa_o$ are time-independent.

\subsubsection{(Strongly) Isolated Horizons}
\label{s:IH}
An isolated horizon (IH) is a WIH on which the whole extrinsic
geometry is time-invariant: $[{\delta}_\ell, \hat{\nabla}_a]=0$.
This condition can be characterized\cite{AshBeeLew02,GouJar06} 
as ${\delta}_\ell \Theta^{(k)}=0$,
that leads to the geometric constraint
\bea
\label{e:IH_characterization}
\kappa^{(\ell)} \Theta^{(k)}_{ab}=
   \frac{1}{2} \left( {}^2\!D_a \Omega^{(\ell)}_b + {}^2\!D_b \Omega^{(\ell)}_a \right) 
+ \Omega^{(\ell)}_a \Omega^{(\ell)}_b 
   - \frac{1}{2}  {}^2\!R_{ab} 
   + 4\pi \left({q^c}_a {q^d}_b T_{cd} - \frac{T}{2} q_{ab}\right) 
\eea
With Eq. (\ref{e:Theta_zero}), this fixes completely the
second fundamental form ${\cal K}^c_{ab}$.
Free data of an IH, $(q_{ab}|_{{\mathcal S}_t}, \Omega^{(\ell)}_a|_{{\mathcal S}_t}, 
\kappa^{(\ell)}|_{\Hor}=\kappa_o)$,
are time independent.
Their geometric (gauge-invariant) content can be encoded 
in the pair\footnote{Note the relation with the  complex scalar 
${\cal K}$ in Refs. \refcite{PenRin84,OweBriChe11}.}:
$({}^2\!R,\mathrm{Im} \Psi_2)$.
On the one hand, ${}^2\!R$ accounts for the gauge-invariant part of $q_{ab}$.
Regarding  $\Omega^{(\ell)}_a$,  from 
$d\omega^{(\ell)} = 2 \mathrm{Im} \Psi_2{}^2\!\epsilon$ and 
$\kappa^{(\ell)}=\mathrm{const}$, it follows 
$d\Omega^{(\ell)} = 2 \mathrm{Im} \Psi_2{}^2\!\epsilon$. On a sphere ${\cal S}_t$ we can write
$\Omega^{(\ell)}_a = \Omega^{\mathrm{div-free}}_a + \Omega^{\mathrm{exact}}_a$, 
so that $\Omega^{\mathrm{exact}}_a={}^2\!D_a g$ is  gauge-dependent [cf. 
(\ref{e:rescaling})]. From $d\Omega^{\mathrm{div-free}}_a=2 \mathrm{Im} \Psi_2$,
the gauge-invariant part of $\Omega^{(\ell)}_a$ is encoded in
$\mathrm{Im} \Psi_2$.

\medskip
\noi{\em IH multipoles of axially symmetric horizons}. 
On an axially symmetric  IH, the gauge-invariant part of the 
geometry, $({}^2\!R,\mathrm{Im} \Psi_2)$, can be described decomposed
onto spherical harmonics.
On an axially symmetric section ${\cal S}_t$ of ${\cal H}$,
a coordinate system can be 
{\em canonically} constructed\cite{AshEngPaw04,SchKriBey06}, such that 
[with $A_{\mathcal H}=4\pi (R_{\mathcal H})^2$]
\bea
q_{ab}dx^a \otimes dx^b = 
(R_{\mathcal H})^2\left( F^{-1} \mathrm{sin}^2\theta d\theta\otimes d\theta +
F d\phi\otimes d\phi \right) \ .
\eea
  In particular, $dA=(R_{\mathcal H})^2\mathrm{sin}\theta d\theta d\phi$ 
({\em round sphere}
area element). We can then use standard spherical harmonics
$Y_{\ell m}(\theta)$, with $m=0$ in this axisymmetric case  
\bea
\int_{{\mathcal S}_t}  Y_{\ell 0}(\theta)Y_{\ell' 0}(\theta)d^2A = (R_{\mathcal
  H})^2 \delta_{\ell \ell'} \ ,
\eea
to define the {\em IH geometric multipoles}\cite{AshEngPaw04} $I_n$ and 
$L_n$
\bea
\label{e:multipoles}
I_n &=& \frac{1}{4} \int_{{\mathcal S}_t} {}^2\!R \; Y_{n0}(\theta) \;d^2A  \nn\\
L_n &=& - \int_{{\mathcal S}_t} \mathrm{Im}\Psi_2 \; Y_{n0}(\theta)
  \;d^2A  \ \ .
\eea
Then, {\em Mass} $M_n$ and {\em Angular Momentum} $J_n$ multipoles are 
defined\cite{AshEngPaw04,SchKriBey06,Jasiulek:2009zf} by adequate
dimensional rescalings of $I_n$ and $L_n$.

\subsubsection{Gauge freedom on a NEH: non-uniqueness of the foliation}
\label{e:NEH_gaugefreedom}
Before proceeding to the dynamical case, we underline 
the existence of a fundamental gauge freedom in the equilibrium (null) case:
{\em any foliation $\{ {\cal S}_t\}$ of a NEH ${\cal H}$
provides a foliation of ${\cal H}$ by marginally trapped surfaces}. 
This is equivalent to the rescaling freedom of the null normal
$\ell^a\to f\ell^a$. Therefore, the amount of gauge freedom in the equilibrium case
is encoded in one arbitrary 
function  $f$ on ${\cal S}_t$.

Note that in this equilibrium horizon context, the relevant 
spacetime geometric object (the hypersurface
${\cal H}$) is {\em unique}, 
whereas the gauge-freedom enters in its evolution description due to
the  {\em non-uniqueness} of its possible foliation by MOTS.

\subsection{Dynamical case}

\subsubsection{{\em Existence} 
and {\em foliation uniqueness} results}
\label{s:DH_fund_results}
Let us introduce two fundamental results following 
from the application of geometric analysis techniques to the
study of dynamical trapping horizons.

\medskip

{\bf Property 1} {\em (Dynamical horizon foliation uniqueness\cite{AshGal05})}.  
{\em Given a dynamical FOTH
$\Hor$, the foliation by marginally 
trapped surfaces is  unique.}

 This first result
 identifies an important {\em rigidity property} of DHs:
the uniqueness of its evolution description.
This is in contrast with the equilibrium null case, with its
freedom in the choice of the foliation.
In particular, on a  dynamical FOTH the evolution vector
is completely determined: $h^a$ unique
up to {\em time} reparametrization.

\medskip

{\bf Property 2}  {\em (Existence of DHs\cite{AndMarSim05,AndMarSim07})}.
%{\scriptsize\orange{[Andersson, Mars \& Simon 05, 07]}}}
{\em Given  a marginally trapped surface $\Sp_0$  satisfying
an appropriate stability condition on a Cauchy hypersurface
$\Sigma$, to each 3+1 spacetime foliation $(\Sigma_t)_{t\in\mathbb{R}}$ it
corresponds a unique dynamical FOTHs $\Hor$
containing  $\Sp_0$ and sliced by marginally trapped surfaces
$\{ \Sp_t \}$ such that
$\Sp_t\subset\Sigma_t$.}

 This second result addresses the Initial Value Problem of DHs,
in particular the existence of an evolution for a given MOTS into a dynamical FOTH.
The result requires a stability condition (namely, ${\cal S}_0$
is required to be {\em stably 
outermost}\cite{AndMarSim05,AndMarSim07,Racz:2008tf,Jaramillo:2011pg}), so that
the  sign of the variation of $\theta^{(\ell)}$ in the inward (outward)
direction is under control. This is essentially 
the {\em outer} condition\cite{Hay94} in the FOTH characterization.

\begin{figure}[]
\begin{center}
%\vglue-1.0cm
\includegraphics[angle=90,width=8.0cm,clip=true,angle=-90]{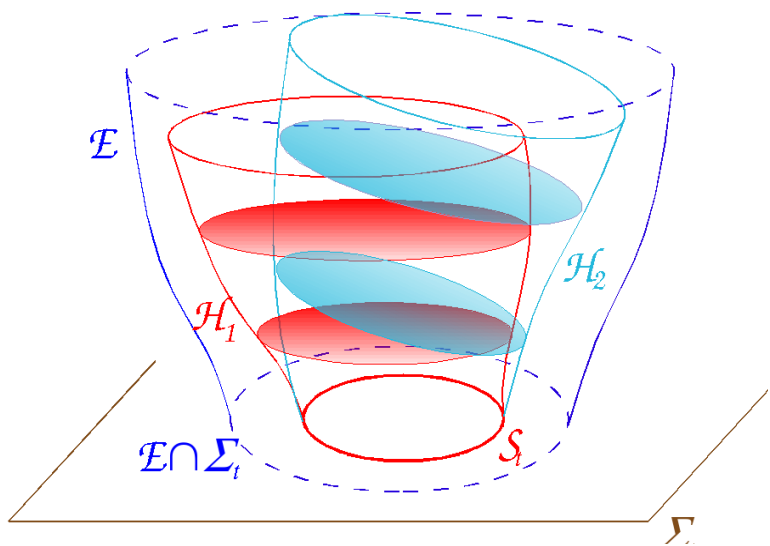}
\end{center}
%\vglue-0.75cm
\caption{
Illustration of the DH non-uniqueness. Dynamical horizons 
${\cal H}_1$ and ${\cal H}_2$ represent 
evolutions from a given initial MOTS corresponding to different spacetime 3+1 slicings. 
}
\label{fig:non_uniqueness}
\end{figure}

\subsubsection{'Gauge' freedom: Non-Uniqueness of Dynamical Horizons}
The evolution of 
an AH into a DH is non-unique, as a consequence of combining Properties 1 and 2
above.
Let us consider an initial AH ${\cal
  S}_0\subset \Sigma_0$ and two different 3+1 slicings
$\{\Sigma_{t_1}\}$ and $\{\Sigma_{t_2}\}$, compatible with 
$\Sigma_0$. From Property 2 there exist DHs ${\cal H}_1 = \bigcup_{t_1}
      {\cal S}_{t_1}$ and ${\cal H}_2 =\bigcup_{t_2} {\cal S}_{t_2} $,
      with ${\cal S}_{t_1} = {\cal H}_1 \cap \Sigma_{t_1}$ and ${\cal
        S}_{t_2} = {\cal H}_2 \cap \Sigma_{t_2}$ marginally trapped
      surfaces.  Let us consider now the sections of ${\cal H}_1$ by
      $\{\Sigma_{t_2}\}$, i.e.  ${\cal S}'_{t_2} = {\cal H}_1 \cap
      \Sigma_{t_2}$, so that ${\cal H}_1 = \bigcup_{t_2} {\cal
        S}'_{t_2}$. In the generic case, slicings $\{{\cal
        S}'_{t_2}\}$ and $\{ {\cal S}_{t_1} \}$ of ${\cal H}_1$ are
      different (one can consider a deformation of  
the slicing $\{\Sigma_{t_2}\}$, if needed).  Therefore,
      from the foliation uniqueness of Property 1, sections ${\cal S}'_{t_2}$ 
      cannot be marginally trapped
      surfaces. It follows then that ${\cal H}_1$ and ${\cal H}_2$
      are different as hypersurfaces in ${\cal M}$: if ${\cal H}_1={\cal H}_2$,
      sections ${\cal S}_{t_2}$ (MOTSs) and ${\cal S}'_{t_2}$ (non-MOTSs)
      would coincide by construction, leading to a contradiction.
In addition to this non-uniqueness, DHs {\em interweave} in spacetime
due to the existence of causal constraints\cite{AshGal05}:
 a DH ${\cal H}_1$ cannot lay completely
in the causal past of another DH ${\cal H}_2$ (cf. Fig. \ref{fig:non_uniqueness}).

Comparing  with the discussion in section \ref{e:NEH_gaugefreedom}
on the uniqueness and gauge-freedom issues in the equilibrium case,
we conclude from the previous geometric considerations 
that the dynamical and equilibrium cases contain
the same amount of gauge freedom, namely a function on ${\cal S}$,
although {\em dressed} in a different form.  More specifically, whereas in 
the NEH case there is a fixed horizon, with a  
rescaling freedom ($\ell^a\to f\ell^a$, $f$ function on ${\cal S}_t$),
in the DH case the foliation is fixed, but
a (gauge) freedom appears in the choice of the evolving horizon
(lapse function $N$ on ${\cal S}_t$).
In other words, 
in the dynamical case the choice is among 
{\em distinct} spacetime geometric objects, ${\cal H}_1$
and ${\cal H}_2$, whereas in the equilibrium case the choice
concerns the {\em description} (foliation) of a single
spacetime geometric object ${\cal H}$.

\subsubsection{FOTH characterization}
\label{e:FOTH_charac_prop}
As discussed in \ref{s:trapping_horizons}, a FOTH
with evolution vector $h^a=\ell^a - Ck^a$ is 
characterized by:
i) a {\em trapping horizon condition}: 
$\theta^{(\ell)}=0, \delta_h \theta^{(\ell)}=0$, ii) a future condition 
$\theta^{(k)}<0$, and iii) an outer condition: $\delta_k \theta^{(\ell)}<0$.
These conditions can be made more explicit in terms of the 
variations\cite{BooFai07,Cao:2010vj}
\bea
\label{e:Aell_Bk}
\delta_{\alpha\ell} \theta^{(\ell)} &=& -\alpha(\sigma^{(\ell)}_{ab} {\sigma^{(\ell)}}^{ab}
 - 8\pi T_{ab}\ell^a\ell^b) \ \\
\delta_{\beta k} \theta^{(\ell)} &=& 
\beta\left[-{}^2\!D^c  \Omega^{(\ell)}_c 
+ \Omega^{(\ell)}_c  {\Omega^{(\ell)}}^c -\frac{1}{2}{}^2\!R
+ 8\pi T_{ab}k^a\ell^b\right] 
+ {}^2\!\Delta \beta - 2 \Omega^{(\ell)}_c  {}^2\!D^c\beta \nn \ ,
\eea
with $\alpha$ and $\beta$ functions on ${\cal S}_t$.
Making $\beta=1$, the outer condition writes
\bea
\label{e:outer}
\delta_{k} \theta^{(\ell)} = 
-{}^2\!D^c  \Omega^{(\ell)}_c 
+ \Omega^{(\ell)}_c  {\Omega^{(\ell)}}^c -\frac{1}{2}{}^2\!R
+ 8\pi T_{ab}k^a\ell^b  < 0 ,
\eea
for some $\ell^a$ and $k^a$,
whereas the trapping horizon condition (with $\alpha=1$, $\beta=C$) is
\bea
\label{e:TH_condition}
\delta_h\theta^{(\ell)} = \delta_\ell \theta^{(\ell)} -\delta_{Ck} \theta^{(\ell)}
=\delta_\ell \theta^{(\ell)} -
C\delta_{k} \theta^{(\ell)}
- {}^2\!\Delta C + 2 \Omega^{(\ell)}_c  {}^2\!D^cC=0  \ , 
\eea
that is
 \bea 
\label{e:trapping_horizon2}
- {}^2\!\Delta C + 2 \Omega^{(\ell)}_c  {}^2\!D^cC 
-C\left[-{}^2\!D^c  \Omega^{(\ell)}_c 
+ \Omega^{(\ell)}_c  {\Omega^{(\ell)}}^c -\frac{1}{2}{}^2\!R\right] =
  \sigma^{(\ell)}_{ab} {\sigma^{(\ell)}}^{ab} + 8\pi T_{ab}\tau^a\ell^b  
\eea
This elliptic condition on $C$, in particular through the
application of a maximum principle relying on the
outer condition $\delta_{k} \theta^{(\ell)}<0$, is at the heart of the non-local
behaviour of the worldtube 
$\bigcup_{t\in\mathbb{R}} \Sp_t $ discussed in section \ref{s:equivalence}.

\medskip

{\em Remark on the variation/deformation/stability operator 
$\delta_v\theta^{(\ell)}$}. Before proceeding further, Eq. (\ref{e:Aell_Bk}) 
requires some explanation. In section \ref{s:extrinsic_S}, we have
introduced $\delta_v$ in terms of the Lie derivative on a tensorial
object. However, the expansion $\theta^{(\ell)}$ is not a scalar quantity in the
sense of a point-like (tensorial) field defined on the manifold ${\cal M}$. The
expansion is a quasi-local object whose very definition at a 
point $p\in{\cal M}$ requires the choice of a (portion of a) surface
 ${\cal S}$ passing through $p$. In this
sense, $\delta_{\gamma v}$ (with $\gamma$ a function on ${\cal S}$)
cannot be  in general evaluated as a Lie derivative. 
Consider a displacement of the surface ${\cal S}_t$ by a
vector $\gamma v^a$.  The surface ${\cal S}_{t+\delta t}$ and therefore
$\theta^{(\ell)}|_{t+\delta t}$ depend on the angular dependence of $\gamma$, so that 
$\delta_{\gamma v}\theta^{(\ell)} \neq \gamma\delta_{v}\theta^{(\ell)}$.
The operator $\delta_v$ still satisfies a linear property
for constant linear combinations, $\delta_{a v + b w}\theta^{(\ell)}=
a\delta_v \theta^{(\ell)}+ b \delta_w\theta^{(\ell)}$
($a, b \in \mathbb{R}$), and
the Leibnitz rule, $\delta_v (\gamma \theta^{(\ell)}) = (\delta_v \gamma) \theta^{(\ell)}
+ \gamma \delta_v \theta^{(\ell)}$.
Details about this operator can be found in Refs. 
\refcite{AndMarSim05,BooFai07,Cao:2010vj}\footnote{See also the treatment 
in terms of Lie derivatives 
in the {\em double null foliations} treatment 
in  Refs. \refcite{Hay94,Gou05}.}.
Here we rather exploit a {\em practical trick} for the evaluation of
$\delta_{\gamma v}\theta^{(\ell)}$, based on the remark  that given the vector $v^a$ 
normal to ${\cal S}$, and {\em not} multiplied by a function on ${\cal S}$, 
it still holds formally $\delta_{v}\theta^{(\ell)} = {\cal L}_v  \theta^{(\ell)} $.
Then, we can evaluate $\delta_{\gamma v}\theta^{(\ell)}$ as  
$\delta_{\gamma v}\theta^{(\ell)}= 
\delta_{\tilde{v}}\theta^{(\ell)}= {\cal L}_{\tilde{v}}  \theta^{(\ell)}$, with 
$\tilde{v}^a=\gamma v^a $. In particular, the application of this strategy to
the second line of (\ref{e:Aell_Bk}) goes as follows.
We write  $\tilde{k}^a = \beta k^a$ and calculate $\delta_{\tilde{k}} \theta^{(\ell)}$
through a Lie derivative evaluation. This results in 
\bea
\label{e:delta_tildek}
\delta_{\tilde{k}} \theta^{(\ell)}= 
(-\tilde{k}^c\ell_c)\left[
 {}^2\!D^c  \Omega^{(\tilde{k})}_c 
 + \Omega^{(\tilde{k})}_c  {\Omega^{(\tilde{k})}}^c -\frac{1}{2}{}^2\!R \right]
 + 8\pi T_{ab}\tilde{k}^a\ell^b \ .
\eea
Using
$(-\tilde{k}^c\ell_c)=\beta$, $\Omega^{(\tilde{k})}_a = \Omega^{(k)}_a 
+{}^2\!D_a\mathrm{ln}\beta$ and $\Omega^{(k)}_a = -\Omega^{(\ell)}_a$ the expression
for  $\delta_{\tilde{k}} \theta^{(\ell)}$ in (\ref{e:Aell_Bk}) follows  
(cf. footnote \ref{f:delta_Omega}).

\subsubsection{Generic properties of dynamical FOTHs}
\label{s:DH_eneric_properties}

We review some generic properties of dynamical trapping 
horizons\cite{Hay94,AshKri02,AshKri03,AshKri04,BooFai07}. 

\medskip

\noi i) {\em Topology Law}:  under the dominant energy condition,
sections ${\cal S}_t$ are topological spheres. This can be shown by
integrating $\delta_k \theta^{(\ell)}<0$ on ${\cal S}_t$. 
Under the assumed energy  condition, the 
%{\small {\em Hint}: Integrate $\delta_k \theta^{(\ell)}$ condition 
%on ${\cal S}_t \to$ positive 
Euler characteristic $\chi$
\be
\label{e:Euler}
\chi = \frac{1}{4\pi}\int_{\cal S}{}^2\!R \; {}^2\!\epsilon=
\frac{1}{2\pi}\int_{\cal S}\left(-\delta_k \theta^{(\ell)}  
+ \Omega^{(\ell)}_c  {\Omega^{(\ell)}}^c
+ 8\pi T_{ab}k^a\ell^b\right)  {}^2\!\epsilon \nn \ ,
\ee
is positive and, being ${\cal S}_t$ a closed 2-surface, its
spherical topology follows.

\noi ii) {\em Signature law}: under the null energy condition, 
${\cal H}$ is completely partitioned into null worldtube sections 
(where $\delta_\ell \theta^{(\ell)}=0$) and spacelike  worldtube
sections (where $\delta_\ell \theta^{(\ell)}\neq 0$ at least on a point). 
Applying a  maximum principle  to the trapping
horizon constraint condition, Eq. 
(\ref{e:TH_condition}), it follows that either 
$C=\mathrm{const}\geq 0$, or $C$ is a function $C>0$ everywhere on ${\cal S}$ (cf. discussion
in \ref{s:equivalence}). 

\noi iii) {\em Area law}: under the null energy condition, if 
$\delta_\ell \theta^{(\ell)}\neq 0$ somewhere on ${\cal S}_t$,
the area grows locally everywhere
on ${\cal S}_t$. Otherwise the area in constant along the evolution. 
This follows from
applying the future condition, $\theta^{(k)}<0$, and
 the signature law to $\delta_{h} {}^2\!\sqrt{q}
= -C\theta^{(k)} \sqrt{q}$ [cf. Eq. (\ref{e:area_change})].

\noi iv) {\em Preferred choice of null tetrad on a DH}. 
According to the {\em foliation uniqueness} and {\em existence} results discussed in 
\ref{s:DH_fund_results}, there is a unique evolution 
vector $h^a$ tangent to $\Hor$ and orthogonal
to ${\cal S}_t$, such that $h^a$ transports ${\cal S}_t\in\Sigma_t $ 
onto ${\cal S}_{t+\delta t}\in \Sigma_{t+\delta t}$: that is,
${\delta}_h t =1$, for a given function $t$ defining a 3+1 spacetime foliation
$\{\Sigma_t\}$. Denoting the unit timelike
normal to $\Sigma_t$ by $n^a$, the lapse function by $N$,
i.e. $n_a=-N\nabla_at$, and the normal
to ${\cal S}_t$ tangent to $\Sigma_t$ by $s^a$, we can write
on the horizon ${\cal H}_N$
\bea
\label{e:h_N_b}
h^a = N n^a + b s^a \ ,
\eea
for some $b$ fixed from $N$ and $C$ in (\ref{e:h_m}), as
$2C=(b-N)(b-N)$. 
The expression of the evolution vector as $h^a=\ell^a - Ck^a$ 
[cf. Eq. (\ref{e:h_m})]
links the scaling of $\ell^a$ and $k^a$ to that of $h^a$. 
In particular, $\ell^a$ is singled out as the only null normal to
${\cal S}_t$ such that
$h^a\to\ell^a$ as the trapping horizon is driven to stationarity ($C\to
0\Leftrightarrow \delta_\ell \theta^{(\ell)}\to 0$). 
Writing generically the null
normals at ${\cal H}_N$ as $\ell^a=f\cdot(n^a + s^a)$ and
$k^a=(n^a-s^a)/(2f)$, Eqs. (\ref{e:h_N_b}) and (\ref{e:h_m})
lead to a preferred scaling of null normals on the  DH ${\cal H}_N$
\bea
\label{e:null_normalization}
\ell_N^a = \frac{N+b}{2}\left(n^a + s^a\right) \,, \ k_N^a = \frac{1}{N+b}\left(n^a - s^a\right) \, \ . 
\eea

\subsubsection{Geometric balance equations}
\label{s:balance}
One of the main  motivations for the development
of quasi-local horizon formalisms is the extension 
of the laws of black hole thermodynamics to dynamical
regimes. This involves in particular finding 
balance equations to control the rate of change of 
{\em physical quantities} on the horizon, in terms
of appropriate fluxes through the hypersurface.
This is an extensive subject
whose review is beyond our scope.
In the spirit of the present  discussion, we restrain ourselves to comment on
the balance equations for two geometric quantities on ${\cal S}_t$:
the area $A=\int_{\cal S}dA=\int_{\cal S} {}^2\!\epsilon$ and the angular momentum $J[\phi]$
in Eq. (\ref{e:Komar_angular_momentum}), for an axial Killing
 (or, more generally, divergence-free)  vector $\phi^a$ . 
That is, we aim at writing 
\be
\label{balance_equations_A_J}
\frac{dA}{dt} = \int_{{\cal S}_t} F^A \;dA \ \ , \ \ 
\frac{dJ[\phi]}{dt} = \int_{{\cal S}_t} F^J \;dA \ ,
\ee
for appropriate area $F^A$ and angular
momentum $F^J$ fluxes, with $d/dt$ associated
to the foliation Lie-transported by $h^a$.
Eventually, one would aim at writing a {\em 1st law of thermodynamics}
by appropriately combining the previous balance equations
\be
\kappa_t\frac{dA}{dt}  + \Omega_t \frac{dJ[\phi]}{dt} = 
\int_{{\cal S}_t} F^E \; dA \ ,
\ee
for some functions $\kappa_t$ and $\Omega_t$ on ${\cal S}_t$, so that
$F^E$ is interpreted as an energy flux\cite{AshKri02,AshKri03,Hay04,Hayward:2004fz,BooFai04,BooFai07,Hayward:2008ti,Wu:2009dm,Wu:2010eu}. 
As a first step towards (\ref{balance_equations_A_J}) 
we write evolution equations for the expansion $\theta^{(h)}$ 
and the form $\Omega^{(\ell)}_a$ along the evolution vector $h^a$.
These equations are given by the projection of some of the 
components of the Einstein equations onto ${\cal H}$. 
Introducing a {\em 4-momentum current density} $p_a=-T_{ab}\tau^b$, with
$\tau^a$ the vector  orthogonal to ${\cal H}$ defined in (\ref{e:h_m}), 
such equations provide three of the components of $p_a$.
The fourth is given by the trapping horizon condition (\ref{e:trapping_horizon2}). 
In brief:

\medskip
\noi i) Evolution element of area\cite{GouJar06b,GouJar08} ($p_a h^a = - T_{ab}\tau^b h^a$):
\bea
\label{e:delta_h_theta_h}
\left(\delta_{h} + \theta^{(h)}\right)\theta^{(h)}  =
	- \kappa^{(h)} \theta^{(h)} 
	+  \sigma^{(h)}_{ab} {\sigma^{(\tau)}}^{ab}
	+ \frac{(\theta^{(h)})^2}{2}   
	- 2 {}^2\!D^a Q_a 
	+ 8\pi T_{ab}\tau^a h^b  
	- \frac{\theta^{(k)}}{8\pi} \delta_hC  \ \ , 
\eea
with $Q_a = \frac{1}{4\pi} \left[ C \Omega^{(\ell)}_a 
- 1/2 {}^2\!D_aC \right]$ and $\kappa^{(h)}= -h^b k^c\nabla_b \ell_c$. 

\noi ii) Evolution normal (rotation) form\cite{Gou05,GouJar08}
$\Omega^{(\ell)}_a$ ($p_b {q^b}_a = -T_{bc}\tau^c{q^b}_a$):
\bea
\label{e:delta_h_Omega}
\left(\delta_{h} + \theta^{(h)}\right) \Omega^{(\ell)}_a 
	 =  {}^2\!D_a \kappa^{(h)}
	- {}^2\!D^c \sigma^{(\tau)}_{ac}
	- {}^2\!D_a\theta^{(h)} 
         + 8\pi {q^b}_aT_{bc}\tau^c 
	- \theta^{(k)} {}^2\!D_a C \ .
\eea

\noi iii) Normal component ($p_a \tau^a = - T_{ab}\tau^b \tau^a$): 
linear combination, using $\tau^a= 2\ell^a - h^a$, of $T_{ab}\tau^a h^b$ 
(area element evolution) and $T_{ab} \tau^a\ell^b$ [trapping horizon constraint
(\ref{e:trapping_horizon2})].

\medskip

In order to derive the evolution equation for $A$, we write
$A=\int_{\cal S}dA =\int_{\cal S}{}^2\!\epsilon$ so that, using the
transport of ${\cal S}_t$ into ${\cal S}_{t+\delta t}$ by $h^a$,
we have $\frac{dA}{dt}= \int_{\cal S}\delta_h (dA)  =\int_{\cal S}\theta^{(h)}dA$
and $\frac{d^2A}{dt^2}=\int_{\cal S}\left(\delta_h\theta^{(h)}+
(\theta^{(h)})^2\right)dA$. From  Eq. (\ref{e:delta_h_theta_h}) it then follows 
\bea
	\frac{d^2 A}{dt^2} + \bar\kappa' \frac{dA}{dt}  = 
	\int_{\Sp_t}  \Big[  
	8\pi T_{ab}\tau^ah^b  
	+ \sigma^{(h)}_{ab} {\sigma^{(\tau)}}^{ab}  
 + \frac{(\theta^{(h)})^2}{2}
+ (\bar{\kappa}'-\kappa') \theta^{(h)}  \; \Big] \, {}^2\!\epsilon \ ,
\eea
where $\kappa' \equiv  \kappa - {\delta}_h\ln C$ and 
$\bar{\kappa}' = \bar{\kappa}(t) \equiv A^{-1} \int_{\Sp_t} \kappa' {}^2\!\epsilon$.
Note that this is a second-order equation for the area\cite{GouJar06b}.
Near equilibrium, the second time derivative as well as higher-order terms
can be neglected leading to the Hawking \& Hartle expression\cite{Hawking:1972hy}
\bea
	\bar\kappa' \frac{dA}{dt}  = 
	\int_{\Sp_t}  \Big[  
	8\pi T_{ab}\ell^a\ell^b  
	+ \sigma^{(\ell)}_{ab}
 {\sigma^{(\ell)}}^{ab} \Big]  dA \ .
	 \nonumber 
\eea
Regarding the evolution equation for $J[\phi]$, we make
use of Eq. (\ref{e:delta_h_Omega}) together with
a divergence-free condition on $\phi^a$ (that relaxes the Killing condition)
and the condition that $\phi^a$ is Lie-dragged
by the evolution vector $h^a$. Then\cite{Gou05,Hayward:2006ss,Hay06b} 
\be 
     \frac{d}{dt} J(\phi) =
     - \int_{\Sp_t} T_{ab}\tau^a\phi^b \; {}^2\!\epsilon
     - \frac{1}{16\pi} \int_{\Sp_t}
     \sigma_{ab}^{(\tau)}\delta_{\phi}q^{ab} \; {}^2\!\epsilon \ ,
\ee
with the second term in the right hand side accounting for a non-Killing 
$\phi^a$.
Interestingly in dynamical (spacelike) horizons ${\cal H}$, the
conditions ${}^2\!D_a\phi^a=0$ and $\delta_h \phi^a$ completely fix\cite{Hayward:2006ss}
the form of the vector $\phi^a$: $\phi^a ={}^2\!\epsilon^{ac}{}^2\!D_b\theta^{(h)}$.

\subsubsection{Open geometric issues and physical remarks}
\label{s:open_issues}
To close this generic section on geometric
aspects of dynamical horizons, 
we list some relevant open 
geometric problems:

\noi i) {\em Canonical choice of dynamical trapping horizon}.
 DHs are highly non-unique in a given black hole spacetime.
A natural question concerns the possibility of making a canonical choice.
There has been some attempts in this direction
based on {\em entropic} 
arguments\cite{GouJar06b,Booth:2010kr,Nielsen:2010wq,Booth:2011qy,}. 
A very interesting avenue
lies on the  recently introduced notion
of the {\em core of the trapped region}\cite{BenSen10} 
(see also J.M.M. Senovilla's contribution).

\noi ii)  {\em Asymptotics of dynamical horizons to the event horizon}.
One would expect DHs to asymptote generically to
the event horizon at late times. 
This is indeed a topic of active 
research\cite{Williams:2007tp,Williams:2009me,Williams:2010pj,Nielsen:2010gm}.

\noi iii) {\em Black hole singularity {\em covering} by
dynamical horizons}. In addition to the asymptotics of DHs to the event horizon,
it is also of interest to assess 
their behaviour at the {\em birth} of the black hole singularity,
in particular their capability to separate ({\em dress}) singularities
from the rest of the spacetime (see section \ref{s:innerhor}). 

\medskip 

{\em DHs as physical surfaces}. 
Dynamical horizons are  objects
with very interesting geometric properties for the study of black hole
spacetimes.
In addition, from a physical perspective it is 
remarkable that they admit a non-trivial thermodynamical description
(cf. A. Nielsen's contribution).
However, it is also important to underline that, if thought as boundaries
of compact physical objects (in the sense we think, say, of the
surface of a neutron star), then they have non-standard physical properties:
\bit
\item[a)] They are  {\em non-unique}. From an 
Initial Value Problem perspective, the question about
the evolution of a given AH is not well-posed,
since it depends on the  3+1 slicing choice 
(such non-uniqueness in evolution is typical of gauge objects).

\item[b)] Dynamical trapping horizons are
{\em superluminal}, something difficult to reconciliate with the 
physical surface of an object.

\item[c)] DHs show a  {\em non-local behaviour}. 
For instance, they grow globally (reacting {\em as a whole}) when 
{\em energy} crosses them at a given local region (even a point). This
is a consequence of their intrinsic  
{\em elliptic}, rather than {\em hyperbolic}, behaviour.

\eit

%%%%%%%%%%%%%%%%%%%%%%%%%%%%%%%%%%%%%%%%%%%%%%%%%%%%%%%%%%%%%%%%%%%%%%%%%%
%%%%%%%%%%%%%%%%%%%%%%%%%%%%%%%%%%%%%%%%%%%%%%%%%%%%%%%%%%%%%%%%%%%%%%%%%%
%%%%%%%%%%%%%%%%%%%%%%%%%%%%%%%%%%%%%%%%%%%%%%%%%%%%%%%%%%%%%%%%%%%%%%%%%%
\section{Black hole spacetimes in an Initial-Boundary Value problem approach}
\label{s:BH_IVP}
%%%%%%%%%%%%%%%%%%%%%%%%%%%%%%%%%%%%%%%%%%%%%%%%%%%%%%%%%%%%%%%%%%%%%%%%%%
%%%%%%%%%%%%%%%%%%%%%%%%%%%%%%%%%%%%%%%%%%%%%%%%%%%%%%%%%%%%%%%%%%%%%%%%%%
%%%%%%%%%%%%%%%%%%%%%%%%%%%%%%%%%%%%%%%%%%%%%%%%%%%%%%%%%%%%%%%%%%%%%%%%%%

In the context of an Initial-Boundary Value Problem approach to 
the construction of spacetimes, dynamical trapping horizons
play a role at two levels: i) first, as an  {\em a priori} ingredient
to be incorporated into a given PDE formulation of Einstein equations, and ii)
as an {\em a posteriori} tool to extract information of the constructed
spacetimes. In this section we address their application as an {\em a priori}
ingredient.

\subsection{The Initial Value Problem in General Relativity: 3+1 formalism}

Our general basic problem is the control\cite{Fri05} of
the qualitative and quantitative aspects of {\em generic} solutions to
Einstein equations in dynamical scenarios involving a black hole spacetime.
The Initial-Boundary Value Problem approach provides
a powerful avenue to it.
Such a strategy is well suited, on the one
hand,  to the use of global analysis and Partial Differential
Equations (PDE) tools for controlling the qualitative aspects of the problem
and, on the other hand, to the employment of numerical techniques 
to assess the quantitative ones.
In particular, we focus here on the Cauchy (and hyperboloidal) Initial Value
Problem.

\subsubsection{Einstein equations: Constraint and Evolution System}
General Relativity is a geometric theory in which not all the fields
constitute physical degrees of freedom (gauge theory), so that
constraints among the fields are present. In the passage from the 
geometric formulation of the theory to an analytic problem in the
form of a specific PDE system, several PDE subsystems enter into 
scene\cite{Jaramillo:2007fx}.
First, the {\em constraint system} is determined by the (Gauss-Codazzi) conditions that
data on a 3-dimensional Riemannian manifold must satisfy to be
considered as initial data on a spacetime slice. The Hamiltonian and momentum
constraints are determined by the  $G_{ab} n^b$ components of the Einstein equation,
where $n^a$ is a unit timelike vector normal to the initial slice. Second,
the {\em evolution system} is built from the rest of Einstein equation,
including possible auxiliary fields. The {\em gauge system} determines 
the dynamical choice of coordinates in the spacetime. Finally, a {\em subsidiary system}
controls the internal consistency of the previous systems.

\subsubsection{3+1 formalism}
\label{s:3+1}
We introduce some notation regarding the 3+1 formalism\cite{Gou07a}.
As in section \ref{s:DH_eneric_properties}, given a
3+1 slicing of spacetime by spacelike hypersurfaces $\{\Sigma_t\}$,
the unit timelike normal to $\Sigma_t$ is denoted by $n^a$ and the lapse function as
$N$, $n_a=-N\nabla_at$, with $t$ the scalar function defining the 3+1 slicing.
The 3+1 evolution vector is denoted by
$t^a = N n^a + \beta^a$,  where $\beta^a$ is the shift vector. 
The induced metric on $\Sigma_t$ is
denoted by  $\gamma_{ab}$, i.e. $\gamma_{ab} = g_{ab} + n_a n_b $. 
We choose the following sign convention for the extrinsic curvature 
of $\Sigma_t$ in ${\cal M}$: 
$K_{ab} = -{\gamma^c}_a\nabla_c n_b= -\frac{1}{2}{\cal L}_n \gamma_{ab}$.
In particular, we can write ${K_{ij}}=\frac{1}{2N} \left( \gamma_{ik} D_j \beta^k
+\gamma_{jk} D_i \beta^k - {\dot\gamma}_{ij} \right)$, where the dot denotes
the derivative ${\cal L}_t$. Indices i,j,k... are used for objects
leaving on $\Sigma_t$.
For concreteness, we focus on a particular 3+1 decomposition of
Einstein equations, namely involving the following 
conformal decomposition ({\em conformal Ansatz}\cite{Lic44}) 
for data $(\gamma_{ij}, K^{ij})$ on $\Sigma_t$
\bea
\label{e:conformal_ansatz}
\gamma_{ij} = \Psi^4 \tilde{\gamma}_{ij} \ \ , \ \
K_{ij} = \Psi^\zeta  \tilde A_{ij} + \frac{1}{3} K \gamma_{ij} \ ,
\eea
for several $\zeta$ choices.
Denoting by $\tilde{D}_i$ the Levi-Civita connection associated with
$\tilde{\gamma}_{ij}$ 
%we express the traceless part of the
%extrinsic curvature as 
%\bea
%\label{e:Atilde}
%\tilde{A}^{ij} = \frac{\Psi^{4-\zeta}}{2N}
%        \left( \tilde{D}^i\beta^j + \tilde{D}^j\beta^i 
%         - \frac{2}{3}  \tilde{D}_k\beta^k \tilde{\gamma}^{ij}
%         + \dot{\tilde\gamma}^{ij}  \right) 
%\eea
and inserting (\ref{e:conformal_ansatz}) into Einstein equations leads to
a coupled elliptic-hyperbolic PDE system on the variables $\Psi$, $\beta^i$,
$N$ and $\tilde{\gamma}_{ab}$. The elliptic part has the form
\bea
\label{e:elliptic_part}
 \tD_k \tD^k \Psi - \frac{{}^3\!{\tilde R}}{8} \, \Psi &=&
 S_{\Psi}[\Psi, N, \beta^i, K, \tilde{\gamma},...] \nonumber \\
    \displaystyle \tD_k \tD^k \beta^i +  \frac{1}{3} \tD^i \tD_k \beta^k 
    + {}^3\!{\tilde R}^i_{\ \, k} \beta^k & = & S_{\beta}[\Psi,
 N, \beta^i, K, \tilde{\gamma},...]  \\
    \displaystyle \tD_k \tD^k N + 2 \tD_k\ln\Psi\, \tD^k N 
    & = & 
    S_{N}[N, \Psi, \beta^i, K, \tilde{\gamma}, \dot{K}, ...]  \nonumber \ ,
\eea
where the equation on $\Psi$ follows from the Hamiltonian constraint, the
equation on $\beta^i$ follows from the momentum constraint and 
the third equation on $N$ follows from
a (gauge) condition imposed on $\dot{K}$. If only solved on an initial slice 
with $\tilde{\gamma}_{ij}$, $ \dot{\tilde\gamma}^{ij}$, $K$ and $\dot{K}$ as
free data, this system
constitutes the {\em Extended Conformal Thin Sandwich} approach to
initial data\cite{Yor99,PfeYor03}. If we solve it during the whole
evolution, together with 
\bea
\label{e:hyperbolic_part}
\frac{\partial^2 {\tilde{\gamma}}^{ij}}{\partial t^2} - \frac{N^2}{\Psi^4}\Delta
{\tilde{\gamma}}^{ij}
- 2 {\cal L}_{\beta}\frac{{\tilde{\gamma}}^{ij}}{\partial t} + {\cal
  L}_{\beta}{\cal L}_{\beta}{\tilde{\gamma}}^{ij} = S_{{\tilde{\gamma}}}^{ij} [N, \Psi, \beta^i, K, \tilde{\gamma}, ...] \ , 
\eea
for $\tilde{\gamma}_{ij}$, it defines a particular constrained evolution 
formalism\cite{BonGouGra04,CorderoCarrion:2008cx,CorderoCarrion:2008nf}.

\subsection{Initial Data: {\em Isolated Horizon} inner boundary conditions}
\label{s:initial_data}
There are two standard approaches to ensure that initial data on a slice $\Sigma_0$  
correspond to a black hole spacetime. The {\em punctures} approach 
exploits the non-trivial topology\cite{Gan75,Gan76} of  $\Sigma_0$, 
whereas the {\em excision} approach removes a 
sphere from the initial slice and  enforces it to be 
inside the black hole region.
In a sense, they both reflect the {\em global} versus {\em quasi-local} 
discussion in section \ref{s:global_local}.
Here we discuss the
use of inner boundary conditions 
derived from the IH formalism, when constructing
initial data of black hole {\em instantaneously in equilibrium} in
an {\em excision approach}.

\subsubsection{Non-Expanding Horizon conditions}
\label{s:ID_NEH}
The NEH condition $\Theta^{(\ell)}_{ab}=0$  in Eq. (\ref{e:Theta_zero}) 
[or (\ref{e:NEH_condition})] provides
three inner boundary conditions for the elliptic system (\ref{e:elliptic_part}).
In particular, they enforce the  excised surface ${\cal S}_0$ to be 
a section of a quasi-local horizon instantaneously in equilibrium.
 
For a given choice of free initial data in system (\ref{e:elliptic_part}),
the geometric NEH  
inner boundary conditions, $\Theta^{(\ell)}_{ab}=0$, must be complemented with two additional 
inner boundary (gauge) conditions.
Denoting by $s^i$ the normal vector to
${\cal S}_t$ tangent to $\Sigma_t$, we write $\beta^i = \beta^\perp s^i + \beta^i_\parallel$, 
with $\beta^\perp = \beta^i s_i$ and $\beta^i_\parallel s_i = 0$. Adapting 
the coordinate system to the horizon (i.e. 
$t^a=\ell^a+\beta^a_\parallel\Leftrightarrow \beta^\perp = N$) supplies a fourth 
gauge condition that, together with the $\theta^{(\ell)}=0$ and
$\sigma_{ab}^{(\ell)}=0$  NEH conditions, 
reads\cite{JarGouMen04,CooPfe04,DaiJarKri04,GouJar06}
\bea
\label{e:NEH_conditions}
&& \tilde{s}^i \tilde{D}_i \Psi +
\tilde{D}_i\tilde{s}^i \Psi+ \Psi^{-1} K_{ij} \tilde{s}^i\tilde{s}^j -
\Psi^3 K = 0  \nn \\
&&\tDS_a \tilde \beta^\parallel_b + \tDS_b \tilde \beta^\parallel_a 
    - (\tDS_c \beta_\parallel^c)\,  \tilde q_{ab}=0  \ \ \ \ , \ \ \   \beta^\perp = N \ ,
\eea
where $\tilde{q}_{ab}= \Psi^4 q_{ab}$ and $\tilde{\beta}^\parallel_a = 
\tilde{q}_{ab}{\beta^\parallel}^b$. A fifth boundary condition, namely for $N$,
can be obtained by choosing a slicing inner boundary condition. 
The (gauge) {\em weakly isolated horizon}
structure can be used in this sense\cite{JarAnsLim07,GouJar06}.

\subsubsection{(Full) Isolated Horizon conditions}
\label{s:ID_IH}
The next geometric  quasi-equilibrium horizon
structure is a (full) IH (cf. sections \ref{s:WIH} and \ref{s:IH}). This involves 
three additional
conditions that cannot be accommodated in system (\ref{e:elliptic_part}) for fixed
free initial data. However, we can revert the argument and employ 
IH conditions to determine improved quasi-equilibrium free initial data
 $\tilde{\gamma}_{ab}$ and $\dot{\tilde{\gamma}}_{ab}$ by solving the full set of
Einstein equations (\ref{e:elliptic_part}) and (\ref{e:hyperbolic_part}) 
under a {\em quasi-equilibrium Ansatz}. Namely, 
we can set $\partial_t \tilde{\gamma}^{ab}$ and 
$\frac{\partial^2 \tilde{\gamma}^{ab}}{\partial t^2}$ in (\ref{e:hyperbolic_part}) 
to prescribed functions
$f_1^{ab}$ and $f_2^{ab}$ and consider the elliptic system formed by 
(\ref{e:elliptic_part}) together with
\bea
- \frac{N^2}{\Psi^4}
\tilde{\Delta}\tilde{\gamma}^{ab} 
 + {\cal
  L}_{\beta}{\cal L}_{\beta}\tilde{\gamma}^{ab} = S_{\tilde{\gamma}}^{ab}
-f_2^{ab} + 2 {\cal L}_{\beta}f_1^{ab} \ .
\eea
This extended elliptic system is solved for ten fields: $(\Psi, \beta^a, N)$ and  the five 
$\tilde{\gamma}^{ab}$. Geometrically, we need to impose four gauge inner conditions, leaving
exactly six inner conditions to be fixed. Remarkably, this fits exactly
the six  IH conditions\cite{Jaramillo:2009cc}
\bea
\Theta^{(\ell)}_{ab}=0 \ \ , \ \ \Theta^{(k)}_{ab}=\Theta^{(k)}_{ab}(\kappa_o, \tilde{q}_{ab}, 
\Omega_a^{(\ell)}) \ \Leftrightarrow \ F_{ab}^{\Theta^{(k)}}(\kappa_o, \Psi, \beta^a, N,\tilde{\gamma}_{ab})=0 \ ,
\eea
where  $F_{ab}^{\Theta^{(k)}}$ is determined by the expression for
$\Theta^{(k)}_{ab}$ in Eq. (\ref{e:IH_characterization}), fixed
up to the value of the constant $\kappa_o$. It is interesting to remark
that this IH prescription\cite{Jaramillo:2009cc} completely fixes 
(up a $\kappa_o$ one-parameter family) 
the  extrinsic curvature tensor 
${\cal K}^c_{ab} = k^c \Theta^{(\ell)}_{ab} + \ell^c \Theta^{(k)}_{ab}$ 
[cf.  Eq. (\ref{e:2nd_fund_form})] of ${\cal S}_0$ as embedded in the 
spacetime ${\cal M}$.

\subsection{Constrained evolutions: {\em Trapping Horizon} inner boundary 
conditions}
The elliptic-hyperbolic system (\ref{e:elliptic_part})-(\ref{e:hyperbolic_part}) 
provides a constrained evolution scheme for the dynamical 
construction of the spacetime.
Adopting a excision approach to black holes,
we need five inner boundary conditions for the elliptic part of
the system.  In principle, dynamical trapping horizon conditions
on the inner boundary worldtube $\Hor=\cup_t {\cal S}_t$
provide a geometric prescription  guaranteeing that $\Hor$
remains in the black hole region. 
However, imposing FOTH conditions 
on ${\cal H}$ can be  {\em too stringent}  in generic evolutions.
The reason is that the constructed worldtube of MOTS ${\cal H}$,
regarded as a hypersurface in spacetime, 
can change signature. This is in conflict with the outer condition 
in \ref{s:trapping_horizons}
(something related to {\em jumps} occurring 
generically\cite{Nielsen:2005af,BooBriGon05,Booth:2007wu} in AH evolutions; see \ref{s:jumps})
so that the resulting PDE system can become ill-posed.
In this context, trapping horizon conditions together with the
requirement of recovering NEH inner conditions 
at the  equilibrium limit, provide an appropriate relaxed set of 
inner boundary conditions\cite{JarGouCor07}.
More specifically, trapping horizon conditions provides two geometric
conditions $\theta^{(\ell)}=0$ and $\delta_{h}\theta^{(\ell)}=0$, whereas
three 
additional gauge conditions guarantee
the recovery of NEH at equilibrium.

As a first step, as in \ref{s:ID_NEH},  we choose a coordinate system 
adapted to the horizon. This means that  spacetime evolution
$t^a$ is tangent to ${\cal H}$.
Decomposing the shift  as $\beta^a = \beta^\perp s^a + \beta_\parallel^a$,
then  $t^a$ is written as $t^a = Nn^a + \beta^a = (Nn^a + bs^a) +  \beta_\parallel^a 
+ (\beta^\perp - b) s^a = h^a +  \beta_\parallel^a + (\beta^\perp - b) s^a$.
Therefore $t^a$ is tangent to ${\cal H}$ if and only if $\beta^\perp = b$.

\medskip
\noi i) {\em Geometric trapping horizon conditions}.
Condition $\theta^{(\ell)}=0$ leads, in terms of the 3+1 quantities
in \ref{s:3+1}, to the expression in the first line 
of Eq. (\ref{e:NEH_conditions}).
Condition  $\delta_{h}\theta^{(\ell)}=0$ in Eq. (\ref{e:trapping_horizon2}),
using the adapted coordinate system
$\beta^\perp = b$, leads to
\bea
\label{e:trapping_hor}
\left[- {}^2\!D_a{}^2\!D^a - 2 L^a {}^2\!D_a + A\right] (\beta^\perp-N) = B (\beta^\perp+N) \ ,
\eea
where
$L_a = K_{ij} s^i q^j_{\ \, a}$, $A =  \frac{1}{2} {}^2\! R - {}^2\!D_a L^a - L_a L^a 
	- 4\pi T_{\mu\nu}(n^\mu+s^\mu)(n^\nu-s^\nu) $,
and $B= \frac{1}{2} \sigma_{ab}^{(\hat{\ell})} \sigma^{(\hat{\ell})ab}
	+ 4\pi T_{ab}(n^a+s^b)(n^b+s^b) $, with $\hat{\ell}^a=n^a+s^a $.

\medskip
\noi ii) {\em Gauge boundary conditions I}. Aiming at recovering 
NEH boundary conditions for $\beta_\parallel^a$,
we first express $\delta_h q_{ab} = \theta^{(h)} q_{ab} + 2 \sigma^{(h)}_{ab} $ 
in adapted coordinates  ($h^a = t^a - \beta_\parallel^a$)
\bea
2\sigma_{ab}^{(h)} = 
 \left(\der{q_{ab}}{t} - \der{}{t}\ln\sqrt{q}\; q_{ab}\right)
	- \left({}^2\!D_a  \beta^\parallel_b + {}^2\!D_b \beta^\parallel_a 
- {}^2\!D_c  \beta_\parallel^c\, q_{ab}\right) \ ,
\eea
Then,  the coordinate choice 
$\partial_t q_{ab} - \partial_t\ln\sqrt{q}\; q_{ab} = 0$ leads to the condition
on $\beta^\parallel_a$
\bea
{}^2\!D_a  \beta^\parallel_b + {}^2\!D_b \beta^\parallel_a 
- {}^2\!D_c  \beta_\parallel^c\, q_{ab} = 
-2 \sigma_{ab}^{(h)} \ ,
\eea
that is completed by using the evolution equation for 
$\sigma_{ab}^{(h)}$ on ${\cal H}$
\bea
\delta_h\sigma^{(h)}_{ab} &=& -{q^d}_a{q^f}_b{C^c}_{def}\ell_c \ell^e
	- C^2 {q^d}_a{q^f}_b{C^c}_{def}k_c k^e \nonumber\\
	&-& 8\pi C \left[ {q^c}_a{q^d}_bT_{cd} - 
        \frac{1}{2} (q^{cd}T_{cd}) q_{ab}\right]
	+ \cdots
\eea

\noi iii)\noi {\em Gauge boundary conditions II}. The slicing condition 
for $N$ is essentially free. However, from Properties 1 and 2 in section 
\ref{s:DH_fund_results},
such a choice is equivalent to choosing 
a dynamical horizon ${\cal H}$.
Since each ${\cal H}$ is a genuine geometric object, this suggests 
the possibility of recasting into {\em geometric} terms
the gauge choice of inner boundary condition for  
$N$, by selecting a trapping horizon ${\cal H}$
satisfying some specific {\em geometric criterion} for ${\cal H}$.
As an example of this,
maximizing the area growth rate $\dot{A}$ of ${\cal H}$ 
leads\cite{GouJar06b,JarGouCor07} to the condition 
$\beta^\perp-N = -\mathrm{const}\cdot\theta^{(\hat{k})}$, with $\hat{k}^a= n^a-s^a$.

%%%%%%%%%%%%%%%%%%%%%%%%%%%%%%%%%%%%%%%%%%%%%%%%%%%%%%%%%%%%%%%%%%%%%%%%%%
%%%%%%%%%%%%%%%%%%%%%%%%%%%%%%%%%%%%%%%%%%%%%%%%%%%%%%%%%%%%%%%%%%%%%%%%%%
%%%%%%%%%%%%%%%%%%%%%%%%%%%%%%%%%%%%%%%%%%%%%%%%%%%%%%%%%%%%%%%%%%%%%%%%%%
\section{{\em A posteriori} analysis of Black Hole spacetimes}
\label{s:BH_aposteriori}
%%%%%%%%%%%%%%%%%%%%%%%%%%%%%%%%%%%%%%%%%%%%%%%%%%%%%%%%%%%%%%%%%%%%%%%%%%
%%%%%%%%%%%%%%%%%%%%%%%%%%%%%%%%%%%%%%%%%%%%%%%%%%%%%%%%%%%%%%%%%%%%%%%%%%
%%%%%%%%%%%%%%%%%%%%%%%%%%%%%%%%%%%%%%%%%%%%%%%%%%%%%%%%%%%%%%%%%%%%%%%%%%

We address here the application of 
dynamical trapping horizons to the {\em a posteriori}
 analysis of spacetimes, their main application in the
Initial Value Problem approach.

\subsection{``Tracking'' the {\em black hole region}: Apparent Horizon
finders }
As discussed in \ref{s:Event_Horizon}, event horizons
cannot be located during the spacetime evolution.
However, in applications such as numerical relativity,  assessing
if  a region of spacetime lays inside the black hole region
can be crucial during the evolution.
Under the assumption of cosmic censorship,
the location of AHs in spatial sections $\Sigma_t$
and the worldtubes constructed by piling them up
(see \ref{s:AH})
are  extremely useful to determine the evolutive properties of the
black hole.
In this sense, {\em apparent horizon finders} 
prove to be extraordinary practical tools. These are algorithms for searching 
surfaces ${\cal S}_t \subset \Sigma_t$ that 
satisfy the MOTS condition $\theta^{(\el)}=0$. There are many approaches to this
problem\cite{Tho07}, but all of them aim at 
solving the condition $D_i s^i - K + K_{ij}s^is^j = 0$. 
For instance, assuming spherical topology, we can characterize the surface
in an adapted (spherical) coordinate system as
$F(r, \theta, \varphi)= r - h(\theta, \varphi)$ with $F=\mathrm{const}$,
so that the normal vector to ${\cal S}_t$ is given by
$s_i=\frac{1}{\sqrt{D^iF\cdot D_iF}}D_iF$ with 
$D_iF=(1, -\partial_\theta h,  -\partial_\varphi h)$ in the spherical coordinate
system. The MOTS condition becomes then a non-linear elliptic equation
on $h$ that can be solved very efficiently.

\subsubsection{Understanding {\em apparent horizon} jumps}
\label{s:jumps}
\begin{figure}[t!]
\begin{center}
\vglue-1.0cm
\includegraphics[angle=90,width=7.0cm,clip=true,angle=-90]{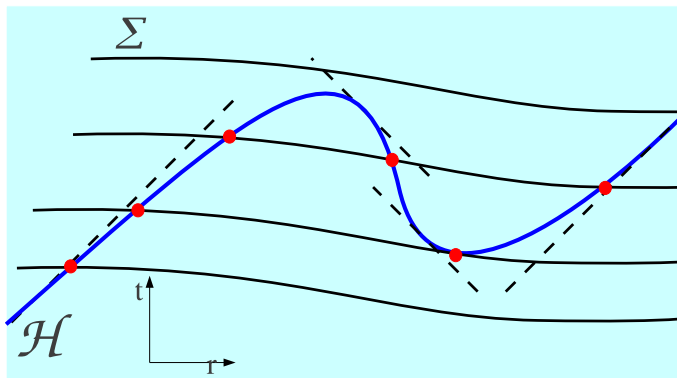}
\end{center}
\vglue-0.75cm
\caption{Illustration
of AH jumps as multiple cuts of a single spacetime
MOTS-worltube ${\cal H}$. In particular, timelike sections of ${\cal H}$ produce
jumps (null hypersurfaces are represented with $45^o$).
  }
\label{fig:jumps}
\end{figure}

Non-continuous {\em jumps} of AHs occur generically in
3+1 black hole evolutions.
The dynamical trapping horizon framework sheds 
light\cite{BooBriGon05,Booth:2007wu,Jaramillo:2009zz} on these
{\em AH jumps}, suggesting a spacetime
picture where the jumps are understood as multiple
spatial cuts of a single underlying spacetime MOTS 
worldtube. Jumps are associated with the change of metric type 
of the horizon hypersurface (see Fig. \ref{fig:jumps}). 
This is particularly dramatic in binary black hole simulations, where
at a given time $t$ the two individual non-connected horizons
{\em jump} to a common one. A specific  prediction of the dynamical
horizon picture is that new (common) horizons form
in {\em pairs}\cite{SchKriBey06,Jaramillo:2009zz}: the 
outermost (apparent) horizon growing in area and a {\em dual}
inner one whose area decrease in the time $t$. Apart from providing
a better understanding of the underlying geometry of the trapped
region, this spacetime
picture can be of use in the study of flows 
interpolating between a given MOTS and the eventual
event horizon, something of potential interest for studies
of the Penrose inequality (see below \ref{s:diagnosis}).

\subsection{Horizon analysis parameters}
Assigning parameters to (individual) black holes
can offer crucial insight into the dynamical evolution.
These can be physical parameters like the mass or the 
angular momentum, or diagnosis parameters informing
of relevant dynamical properties.
Given the generic absence of background rigid structures,
first-principles parameters are often out of reach and one must
follow non-rigorous or {\em pragmatic} approaches.

\subsubsection{Mass and Angular Momentum. IH and DH multipoles}
In our discussion we have avoided entering into first-principles
physical issues, stressing rather the geometric properties of
dynamical trapping horizons and their applications.
However,  mass and angular momentum estimates
for individual black holes, either fundamental or effective,  are extremely important 
in the modeling of astrophysical systems involving matter or binary systems.
The problem has two aspects.
First, one must identify a surface to be associated with the
black hole boundary.
Discussion in section \ref{s:global_local} shows that
this is a delicate question.
In any case,  AHs provide surfaces  ${\cal S}_t\in\Sigma_t$ 
tracking the black hole region, that can be
employed as preferred choices for pragmatic estimations.
The second problem refers to the ambiguities in the quasi-local
characterization of the gravitational field mass and angular momentum in General
Relativity\cite{Szaba09,JarGou10}. Regarding the angular momentum, 
the Komar expression (\ref{e:Komar_angular_momentum}) characterizes 
appropriately the axisymmetric case. Effective 
prescriptions\cite{DreKriSho03,CooWhi07,Kor07}
exist for generic horizons.
Regarding the  mass, the irreducible mass $M_{\mathrm{irred}}$
$A=16\pi M^2_{\mathrm{irred}}$ provides a purely geometric estimation in terms
of the area.
Its physical interpretation as the portion of the black hole
mass that cannot be  extracted by a Penrose process, together with
its equivalence with the Hawking energy, 
$M_{\mathrm{Hawking}}=\sqrt{A/(16\pi)}(1 +1/(8\pi)\oint \theta^{(\ell)}\theta^{(k)}dA)$
for MOTSs, makes it useful in numerical applications and 
in the thermodynamical treatments\cite{Hay04,Hayward:2004fz}.
Given $A$ and $J$ 
one can also consider\cite{AshKri04} the 
Christodoulou expression for the Kerr mass
\be
M_{_{\mathrm{Chris}}} = \left(\frac{A}{16\pi} 
+ \frac{4\pi J^2}{A}\right)^{\frac{1}{2}}  \ .
\ee
There are many prescriptions
for the {\em quasi-local mass}\cite{Szaba09,JarGou10}. It is therefore
crucial to choose and keep consistently a prescription 
when comparing different solutions. In this latter sense, 
 the {\em mass} and {\em angular momentum} horizon 
geometric multipoles $I_n$ and $L_n$ in (\ref{e:multipoles})
offer a useful and refined diagnosis tool in 
numerical studies\cite{SchKriBey06,Vasset:2009pf}.

\subsubsection{Useful diagnosis parameters}
\label{s:diagnosis}
Insight into the geometric properties of MOTS worldtubes 
leads to useful diagnosis parameters for monitoring 
dynamical evolutions. Geometric black hole
inequalities provide a particular avenue.
 In particular, the conjectured Penrose's inequality 
$A\leq 16 \pi M_{_{\rm ADM}}^2$ for asymptotically flat spacetimes
provides a bound to the AH area 
(strictly speaking, the bound is on the area of a minimal 
surface enclosing the AH).
A violation of $\epsilon_{_{\mathrm{Penrose}}}\equiv A/( 16 \pi M_{_{\rm ADM}}^2)\leq 1$ 
{\em indicates} a more exterior MOTS.
In the axially symmetric case this can be refined in terms 
of a so-called\cite{DaiLouTak02,JarVasAns07} {\em Dain number} 
\bea
\label{e:Dain_number}
\epsilon_{_{\mathrm{Dain}}} \equiv \frac{A}{ 8\pi \left(M_{_{\rm ADM}}^2 + 
\sqrt{ M_{_{\rm ADM}}^4 - J^2}\;\right )} \leq 1\ \ .
\eea
Moreover, the rigidity part of the conjecture
provides an extremely simple characterization of Kerr as 
satisfying $\epsilon_{_{\mathrm{Dain}}}=1$. In the same spirit,
the geometric inequality\cite{Dai06a} $J\leq M_{_{\rm ADM}}^2$ 
provides a characterization of (sub)extremality of black holes.
However, these inequalities  involve
total quantities such as the ADM mass. It is remarkable that the 
dynamical horizon structure (actually the 
{\em outer} trapping horizon condition) provides exactly the needed
conditions to prove the quasi-local inequality\cite{Hennig:2008zy,Dain:2011pi,Jaramillo:2011pg}
\bea
\label{e:A_J_ineq}
A\geq 8\pi |J| \ ,
\eea
in generic spacetimes with matter satisfying
the dominant energy condition. 
The validity of the area-angular momentum
inequality (\ref{e:A_J_ineq}) is equivalent to the non-negativity
of the surface gravity $\kappa$ of isolated and dynamical horizons\cite{AshKri04},
supporting the
internal consistency of their first law of black hole thermodynamics.
Inequality (\ref{e:A_J_ineq}) provides a quasi-local characterization of black hole 
(sub)extremality, that is directly related to changes in the horizon metric 
type\cite{Booth:2007wu} and jumps discussed in \ref{s:jumps}.
This is also the context of the 
{\em Booth \& Fairhurst extremality parameter}\cite{Booth:2007wu,Boo07}
\bea
e\equiv 1 + \frac{1}{4\pi}
\int_{\cal S} dA\;\delta_k\theta^{(\ell)} \leq 1 \ .
\eea

\subsection{{\em Heuristic} and effective approaches in {\em a posteriori} 
spacetime analysis}
Hitherto we have discussed
analysis tools to be applied in  numerically constructed
spacetimes, but related to sound geometric
structures. However, when developing a qualitative understanding
of the underlying dynamics, involving e.g. a comparison
with Newtonian or Special Relativity scenarios, the available geometric notions
are often not enough. This is manifest in 
astrophysical contexts requiring estimations for 
linear, orbital angular momentum or binding energies. 
In some cases, a choice must be done between saying nothing at all or
rather adopting a {\em heuristic} approach.

An example of the latter is the following {\em heuristic} 
proposal\cite{KriLouZlo07} for a
quasi-local  black hole linear momentum.
Given a vector $\xi^a$ transverse to a MOTS ${\cal S}$, applying on ${\cal S}$
the linear momentum ADM prescription 
at spatial infinity leads to
\bea
\label{e:P_Krishnan}
P(\xi) = \frac{1}{8\pi}\int_{{\cal S}_t} \left(K_{ab} -K \gamma_{ab}\right)
\xi^a s^b \; {}^2\!\epsilon  \ .
\eea
In spite of its {\em ad hoc} nature, this quantity has been successfully
applied in the analysis\cite{KriLouZlo07} of linear and orbital angular momentum 
in binary black hole orbits and in the recoil dynamics 
of the black hole resulting of asymmetric binary mergers.

\subsection{An effective {\em correlation} approach to the analysis of spacetime dynamics}
\label{s:effective_approach}
The qualitative and quantitative understanding of strong-field
spacetime dynamics represents a challenge in gravitational physics both at
a fundamental level and in applications.
In astrophysical settings a natural strategy consists in
extending to  general relativistic scenarios
the Newtonian {\em celestial mechanics} approach.
This has indeed led to fundamental achievements in the understanding of the 
physics of compact objects.
However, the focus on the properties of individual objects, in particular
in multi-component systems,  
also meets fundamental obstacles in a gravitational 
theory i) without {\em a priori} rigid structures providing canonical structures, 
and ii) with global aspects playing a crucial role. 
The latter encompasses global causal issues and also
the in-built elliptic character of certain objects, both aspects relevant
in the characterization of black holes.
In this context, an approach to spacetime analysis that explicitly
emphasizes  the global/quasi-local properties of the relevant fields, 
at the price of renouncing to a 
detailed tracking of the geometry and {\em trajectories} of small compact regions,
can offer complementary insights to the {\em celestial mechanics} approach. Such
a {\em coarse-grained effective} description is much in
the spirit of the {\em correlation} approach  in the
analysis of complex condensed-matter systems or in quantum/statistical-field theory,
where the functional structure of the (local) dynamical fields is encoded
in the associated {\em n-point correlation functionals}\footnote{N-point 
correlation functions encode the functional structure of the local fields.
A coarse-grained description appear as a truncation to a finite number of
n-point functions.}.  
Such an approach underlines the {\em relational aspects} of the theory, as a 
complementary methodology to the isolation of the dynamical properties a compact parts
of the system. In sum, we can paraphrase the strategy as aiming at {\em a functional
and coarse-grained description of the spacetime geometry, by importing functional tools
for the analysis of condensed matter and quantum/statistical field theory systems.}

\subsubsection{Cross-correlations of geometric quantities at {\em test screens}}
\label{s:cross_correlations}
\begin{figure}[t!]
\begin{center}
\vglue-1.0cm
\includegraphics[angle=90,width=13.0cm,clip=true,angle=-90]{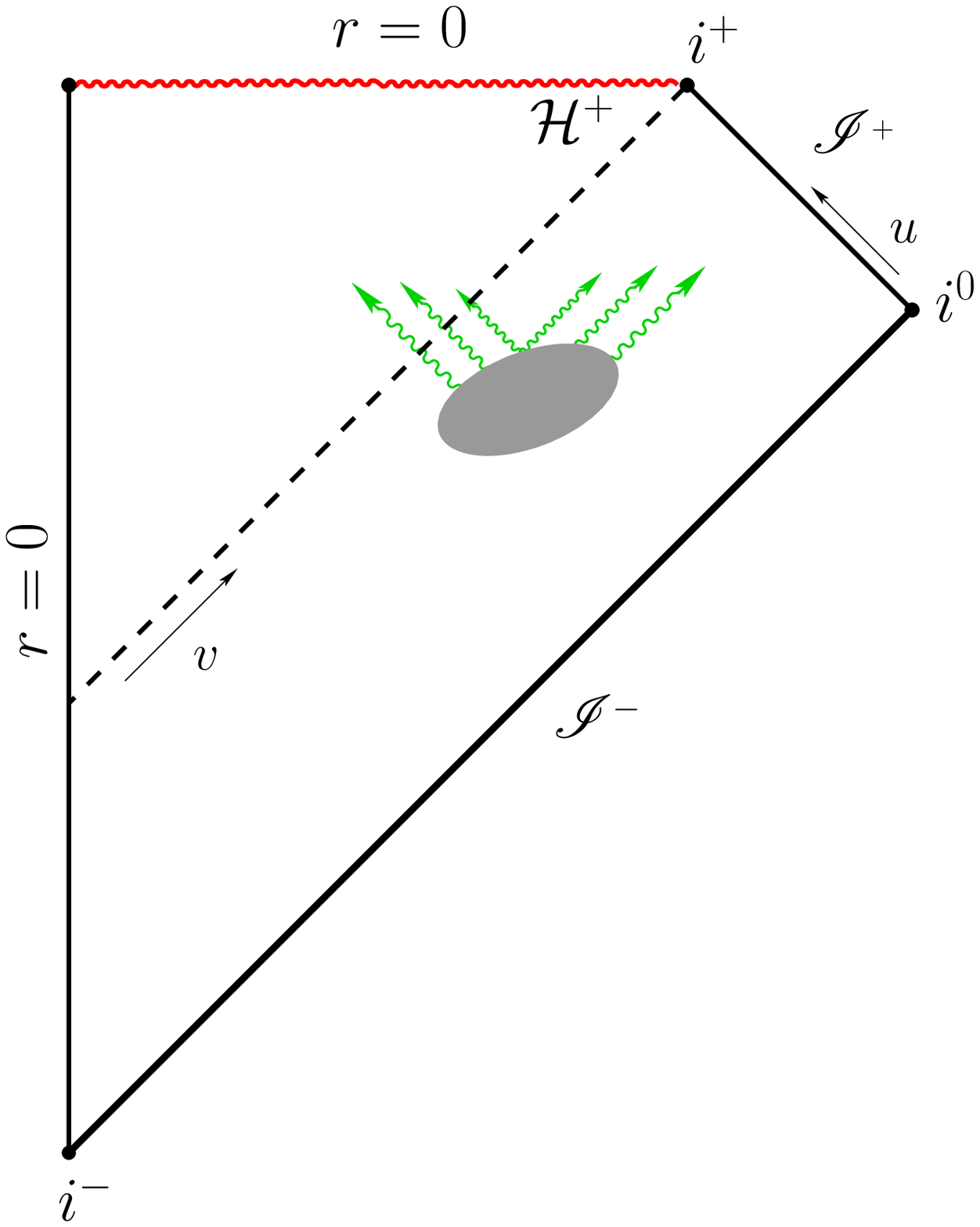}
\end{center}
\vglue-3.75cm
\caption{Carter-Penrose diagram representing a generic (spherically symmetric)
collapse and illustrating 
the {\em cross-correlation} approach to near-horizon gravitational 
dynamics. 
  %The event horizon
  %${\cal H}^+$ and null infinity $\scri^+$ provide spacetime canonical
  %screens on which {\em geometric observables}, respectively
  %accounting for horizon deformations and wave emission, are
  %defined. Their cross-correlation encodes nontrivially information
  %about the bulk spacetime dynamics.
  }
\label{fig:BHscattering}
\end{figure}

The strategy outlined above is admittedly vague.
We sketch now a particular implementation\cite{Jaramillo:2011re,Jaramillo:2011rf} of some of its aspects 
in a {\em cross-correlation} approach to the analysis
of spacetime dynamics.
Aiming at studying the gravitational dynamics in a given spacetime 
region ${\cal R}$, we consider an {\em outer} ${\cal B}_o$ and an {\em inner}  ${\cal B}_i$
hypersurfaces lying in the causal future 
of ${\cal R}$. These hypersurfaces are taken as outer and inner 
boundaries of the bulk spacetime region of interest. 
The geometry of  ${\cal B}_o$ and ${\cal B}_i$  is causally affected by the dynamics 
in ${\cal R}$, so that ${\cal B}_o$ and ${\cal B}_i$ can be understood 
as {\em balloon probes} into the spacetime geometry.
In other words, ${\cal B}_o$ and ${\cal B}_i$ provide {\em test screens} 
(they do not {\em back-react} on the bulk dynamics) on which we can construct
{\em geometric quantities} $h_o$ and $h_i$ to be cross-correlated.
Choosing causally disconnected screens ${\cal B}_o$ and ${\cal B}_i$, 
a non-trivial correlation between $h_o$ and  $h_i$  encodes 
geometric information about the common past region ${\cal R}$.
We can think of this as the reconstruction
of the interaction region from the debris in a {\em scattering} experiment
({\em inverse scattering} picture).
Let us now restrict to the study of near-horizon spacetime dynamics\cite{Jaramillo:2011re,Jaramillo:2011rf}.
In an (asymptotically flat) black hole spacetime setting, 
null infinity $\scri^+$ and the (event) black
hole horizon ${\scre}$ provide {\em canonical} choices
for ${\cal B}_o$  and ${\cal B}_i$, respectively (cf. Fig. \ref{fig:BHscattering}).
Retarded and advanced null coordinates $u$ and $v$ provide
good parameters for  quantities $h_o$ and $h_i$ calculated as
integrals on sections ${\cal S}_u\subset\scri^+$ and  ${\cal S}_v\subset{\cal E}$.
A meaningful notion for the cross-correlation between $h_o(u)$ and $h_i(v)$,
considered as time series, requires the introduction of a (gauge-dependent) mapping 
between $u$ and $v$ at $\scri^+$ and ${\cal E}$. We refer to this point as 
the {\em  time-stretching issue}. 

\subsubsection{Cross-correlations in an Initial Value Problem approach: 
dynamical horizons as canonical inner probe screens}
\label{s:crosscorrelation_IVP}
\begin{figure}[t!]
\begin{center}
\vglue-1.0cm
\includegraphics[angle=90,width=13cm,clip=true,angle=-90]{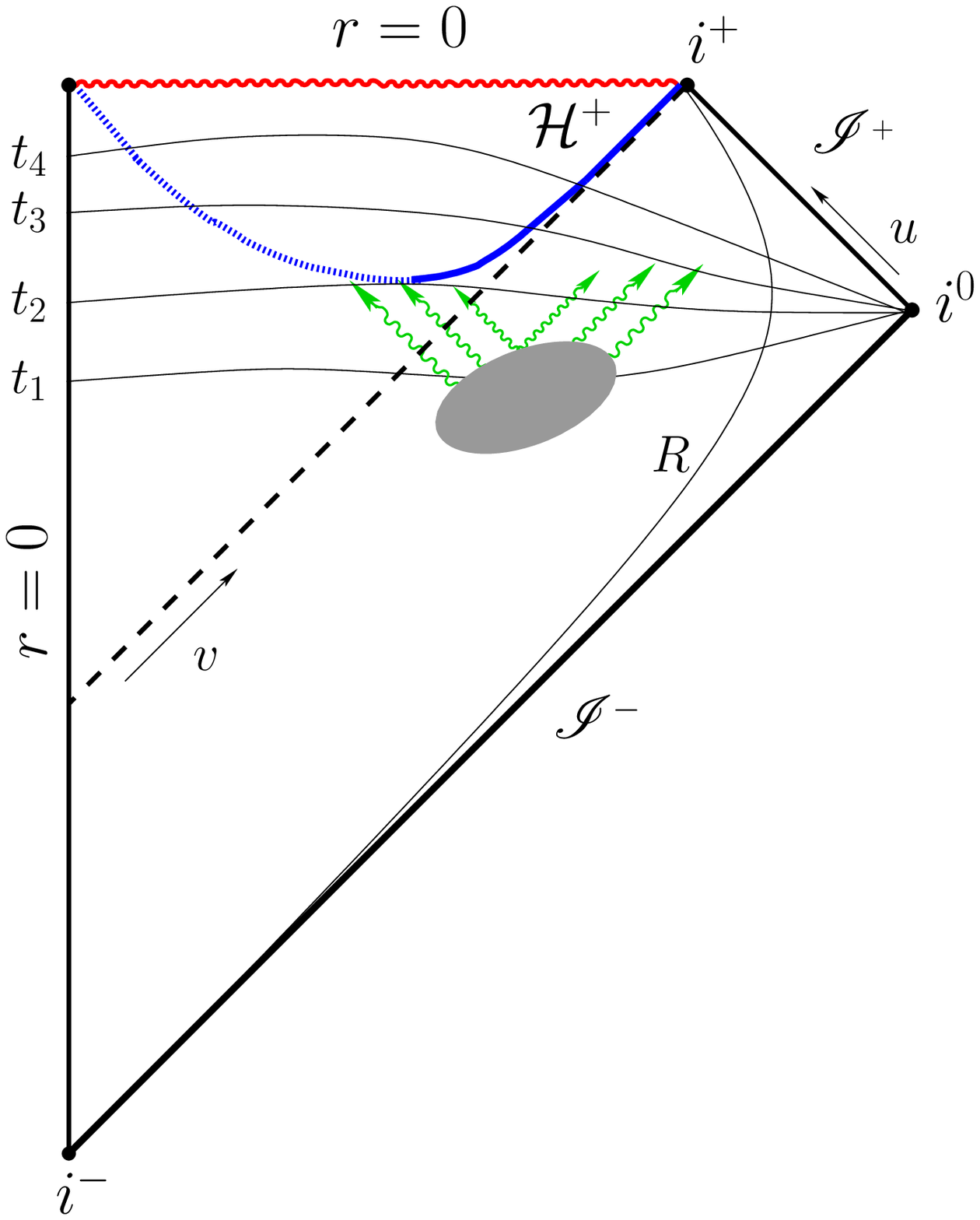}
\end{center}
\vglue-3.75cm
\caption{Carter-Penrose diagram for the {\em cross-correlation} picture in a
  Cauchy IVP approach. 
  %The dynamical horizon ${\cal H}^+$
  %and a large-distance timelike hypersurface ${\cal B}$ provide inner
  %and outer screens. Note that a the dynamical horizon is split in two
  %portions: outer and inner (solid and dashed blue lines,
  %respectively) and that the $3+1$ slicing sets a common time $t$ for
  %cross-correlations.
  }
\label{fig:BHscattering_3+1}
\end{figure}

The adopted Initial Value Problem approach has a direct
impact in the {\em cross-correlation} picture
above.  In particular, the event horizon is not
available during the evolution\footnote{Regarding $\scri^+$,
a pragmatic choice in a Cauchy approach consists
in substituting it by a timelike worldtube of large radii spheres.
However, $\scri^+$ can be kept if using a  
hyperboloidal foliation.}.
Instead, the  (outermost) DH ${\mathcal H}$ fixed by the chosen
3+1 foliation stands as a natural spacetime inner boundary  ${\cal B}_i$.
Although {\em any} hypersurface {\em covering} the black hole
singularity could be envisaged for the present cross-correlation purposes, 
the DH ${\cal H}$ provides a 
natural geometric prescription.
Regarding the {\em time-stretching issue}, 
the time function $t$ defining
the 3+1 spacetime slicing automatically implements a (gauge)
mapping between {\em retarded} and {\em advanced} times $u$ and $v$.
Cross-correlations between geometric quantities at ${\mathcal H}$ and 
$\scri^+$ can then be calculated as standard time-series 
$h_i(t)$ and $h_o(t)$
(cf. Fig.~\ref{fig:BHscattering_3+1}). 
Due to the gauge nature of $t$,  the geometric
information  in quantities $h_i(t)$ and $h_2(t)$ is not encoded in their
{\em local} (arbitrary) time dependence, but rather in the {\em  global} structure of 
successive maxima and minima. The calculation
of cross-correlations must take this into account\cite{Jaramillo:2011re,Jaramillo:2011rf}.
This means, in particular, that quantities to be correlated must be scalars.

\subsubsection{Application to black hole recoil dynamics: towards
DH news functions}
\label{s:recoil_crosscorrelation}
In the context of the study of black hole recoil dynamics after an asymmetric merger,
let us take $h_o(u)$ as the Bondi flux of linear
momentum along a (preferred) direction
\bea
\label{e:flux_Bondi_momentum_u}
\frac{dP^{\mathrm{B}}[\xi]}{du}(u) =
\lim 
\limits_{(u, r \to \infty)}
\frac{r^2}{8 \pi} \oint_{{\cal S}_{u,r}} (\xi^i s_i)  \; |{\cal N}(u)|^2 d\Omega \,,
\ \  \ {\cal N}(u) = 
\int_{-\infty}^u \Psi_4(u') du' \,.
\eea
Here ${\cal N}$ is the news function at $\scri^+$, and $\xi^a$ is a
given spacelike transverse direction to ${\cal S}_{u,r}$, so that $(dP^{\mathrm{B}}/du)[\xi]$
is a scalar.
A natural choice\footnote{An effective curvature vector\cite{JarMacMos11,Rezzolla:2010df} 
constructed 
from the Ricci scalar ${}^2\!R$ on sections ${\cal S}_v$ of ${\cal H}$
provides an intrinsic prescription for 
$h_i(v)$ leading to non-trivial\cite{Jaramillo:2011re} cross-correlations 
with $(dP^{\mathrm{B}}/du)[\xi]$.}
for $h_i(v)$ would be given by the expression
(\ref{e:flux_Bondi_momentum_u}) with  $\Psi_4$ at $\scri^+$
substituted by some $\Psi_0$ at ${\cal H}$. 
A preferred null
tetrad on ${\cal S}_v$ is then needed, something that for DHs is 
 provided by  $\ell^a_N$ and $k^a_N$  in
(\ref{e:null_normalization}). Using them in (\ref{e:Weyl_scalars}), the
preferred Weyl scalar $\Psi^N_0$ is employed to construct
\bea
\label{e:KN}
\tilde{K}^N[\xi](v) =  - \frac{1}{8 \pi}
\oint_{{\cal S}_v} (\xi^is_i) \left|\tilde{\cal N}_N^{(0)}(v)\right|^2 dA \,,
\ \ \hbox{with} \ \ 
\tilde{\cal N}_N^{(0)}(v) 
= \int_{v_0}^v \Psi^N_0(v') dv' \,.
\eea
In spite of the formal similarity between (\ref{e:flux_Bondi_momentum_u})
and (\ref{e:KN}) there is a fundamental difference: whereas
$(dP^{\mathrm{B}}/du)[\xi]$ is an instantaneous flux
through $\scri^+$, this is not true for $\tilde{K}^N[\xi](v)$. The function
 ${\cal N}(u)$  can be written in terms  of geometric quantities
on sections ${\cal S}_u$. This {\em local-in-time} behaviour is a crucial
feature of any valid {\em news function} and it is not shared
by $\tilde{\cal N}_N^{(0)}(v)$. However, it suffices to modify
$\tilde{\cal N}_N^{(0)}(v)$
with terms completing the integrand $\Psi^N_0(v')$ 
to a total differential in time.
Noting ${q^c}_a{q^d}_bC_{lcfd}\ell^l\ell^f = \Psi_0 \overline{m}_a \overline{m}_b +
\overline{\Psi}_0 m_a m_b$, 
inspection of Eq. (\ref{e:null_tidal}) [actually its dynamical version
with $h^a$ instead of $\ell^a$] suggests the identification
of a correct {\em news-like} function at ${\cal H}$ as proportional to the
shear $\sigma^{(h)}_{ab}$ (see also Refs. \refcite{Haywa94b,Haywa03}
for the discussion of the news in quasi-local contexts). In tensorial notation, we write
\bea
\label{e:flux_H}
\frac{dP^{N}}{dv}[\xi](v) =
-\frac{1}{8 \pi}
\oint_{{\cal S}_v} (\xi^i s_i)
\left({\cal N}^{\mathrm{N,g}}_{ab}{\cal N}_{\mathrm{N,g}}^{ab}\right)dA ,
\ \ \hbox{with} \ \  {\cal N}^{\mathrm{N,g}}_{ab} = - \frac{1}{\sqrt{2}} \sigma^{(h)}_{ab} ,
\eea
where the coefficient in ${\cal N}^{\mathrm{N,g}}_{ab}$ guarantees the correct
factor in the leading-term. This $(dP^{N}/dv)[\xi]$ provides  a natural quantity
to be correlated with $(dP^{\mathrm{B}}/du)[\xi]$.
The notation underlines the local character in time
as the {\em flux} of a quantity $P^{N}[\xi]$, but 
no physical meaning is given to the latter. It is worthwhile, though,
to remark the formal similarity of the 
monopolar part of the square of the news ${\cal N}^{\mathrm{N,g}}_{ab}$, i.e
\bea
\label{e:fluxes_E}
\frac{dE^{N}}{dv}(v) = \frac{1}{16 \pi}
\oint_{{\cal S}_v} {\sigma}^{(h)}_{ab}{\sigma}_{(h)}^{ab}dA=
\frac{1}{16 \pi}
\oint_{\cal S} \left[\sigma^{(\ell)}_{ab}{\sigma^{(\ell)}}^{ab}
 -2C\sigma^{(\ell)}_{ab}{\sigma^{(k)}}^{ab} 
+C^2 \sigma^{(k)}_{ab}{\sigma^{(k)}}^{ab}\right]dA
\eea
with the expression of the flux of gravitational 
energy\cite{AshKri02,AshKri03} through a DH, in particular
with its transverse part\cite{Hay04,Hayward:2004fz}.
The identification of ${\sigma}^{(h)}_{ab}$ as a news-like function
suggests a further step,
by introducing a heuristic notion of {\em Bondi-like} 4-momentum 
flux through ${\cal H}$. 
Considering the unit normal $\hat{\tau}^a$ to ${\cal H}$ 
($\hat{\tau}^a=\tau^a/\sqrt{|\tau^b\tau_b|}=
 (\ell^a+Ck^a)/\sqrt{2C}= (bn^a + Ns^a)/\sqrt{2C}$), and 
for a generic spacetime vector $\eta^a$
\bea
\label{e:Bondi_4momentum}
\frac{dP_\tau^N}{dv}[\eta] = -\frac{1}{16 \pi}
\oint_{{\cal S}_v} (\eta^a \hat{\tau}_a) {\sigma}^{(h)}_{ab}{\sigma}_{(h)}^{ab}dA  \ ,
\eea
has formally the expression
of a {\em Bondi-like} 4-momentum\footnote{An alternative expression would follow
by using
in (\ref{e:Bondi_4momentum}), instead of 
$\sigma^{(h)}_{ab}{\sigma^{(h)}}^{ab}$,
the integrand in the DH energy 
flux\cite{AshKri02,AshKri03,Hay04,Hayward:2004fz}, that would also
include the {\em longitudinal} part $\Omega^{(\ell)}_a{\Omega^{(\ell)}}^a$.}.
The flux of energy associated with an Eulerian 
observer $n^a$ would be
\bea
\label{e:energy_tau}
\frac{dE_\tau^{N}}{dv}(v) \equiv \frac{dP_\tau^N}{dv}[n^a]  =
\frac{1}{16 \pi}
\oint_{\cal S} \frac{b}{\sqrt{2C}}\left(\sigma^{(h)}_{ab}{\sigma^{(h)}}^{ab}\right)
dA  \ \,, \nn \\
\eea
where $\frac{b}{\sqrt{2C}}=\sqrt{1+N^2/2C}$. The flux of linear 
momentum for $\xi^a\in T\Sigma_t$ would be
\bea
\label{e:momentum_tau}
\frac{dP_\tau^N}{dv}[\xi] = -\frac{1}{16 \pi}
\oint_{{\cal S}_v} \frac{N}{\sqrt{2C}}(\xi^a s_a) 
\left(\sigma^{(h)}_{ab}{\sigma^{(h)}}^{ab}\right)dA \ .
\eea
Near equilibrium ($C\to0$), we have ${\sigma}^{(h)}_{ab}{\sigma}_{(h)}^{ab}\sim C$ on
DHs [cf. Eq. (\ref{e:area_change})] 
so that expressions (\ref{e:energy_tau}) and (\ref{e:momentum_tau}) are regular
($O(\sqrt{C})$). Integrating (\ref{e:momentum_tau}) in time
would lead to a Bondi-like counterpart\footnote{A related prescription
for a DH linear momentum flux would be given by angular integration
of the appropriate components in the {\em effective gravitational-radiation energy-tensor}
of Ref. \refcite{Hayward:2004fz}.} of the heuristic {\em ADM-like}
linear momentum in (\ref{e:P_Krishnan}).

Before finishing this section, let us mention that the present discussion
on horizon news-like functions can be 
related\cite{Jaramillo:2011rf} to a {\em viscous fluid analogy} for quasi-local 
horizons\cite{Gou05,GouJar06b}. In particular, geometric
{\em decay} and {\em oscillation timescales}  (respectively, $\tau$
and $T$) can be constructed on the
horizon\cite{Jaramillo:2011rf}  from the expansion $\theta^{(h)}$ and
shear $\sigma^{(h)}_{ab}$, respectively related to 
bulk and shear viscosity terms. In the context of black hole
recoil dynamics, this provides an instantaneous geometric
prescription for a {\em slowness parameter}\cite{Price:2011fm} 
$P=T/\tau$ controlling the qualitative aspects of the dynamics.

\subsubsection{The role of the inner horizon in the integration of fluxes along ${\cal H}$}
\label{s:innerhor}
\begin{figure}[t!]
\begin{center}
\vglue-1.0cm
\includegraphics[angle=90,width=13.0cm,clip=true,angle=-90]{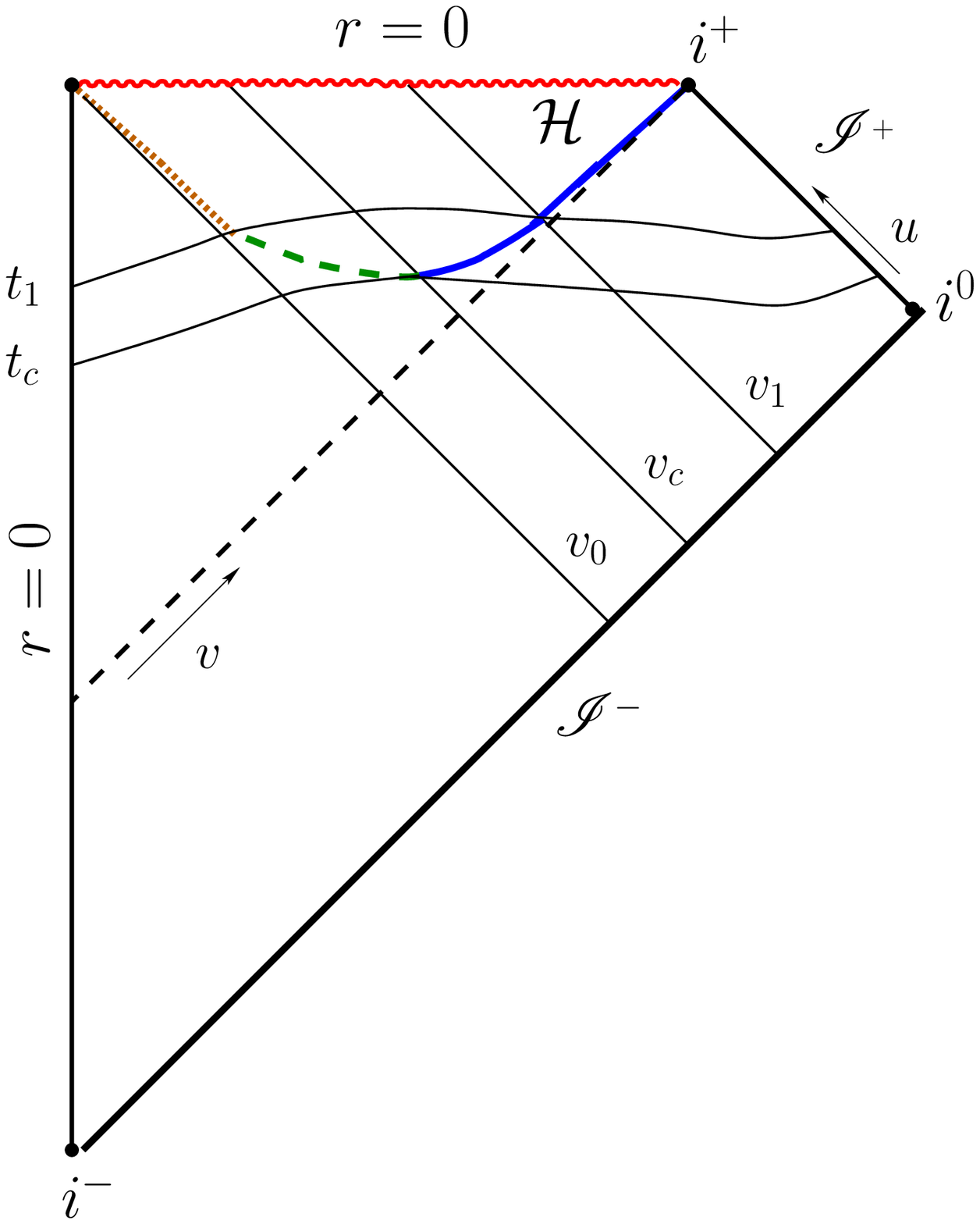}
\end{center}
\vglue-3.75cm
\caption{ Illustration of the splitting of a DH into {\em internal} and 
{\em external} sections by a 3+1 slicing.
  }
\label{fig:inner_hor}
\end{figure}

Flux integrations along ${\cal H}$ require appropriate parametrizations of ${\cal H}$, 
such as an {\em advanced time} $v$. Then,  given the flux
$F_Q(v)$ of a  quantity $Q(v)$, we can write\footnote{The 
coefficient $\mathrm{sign}(C)$,
$+1$ for spacelike ${\cal H}$ and $-1$ for timelike  ${\cal H}$, takes
into account the possible integration of fluxes happening when  
timelike sections of ${\cal H}$ occur; cf. Fig.~\ref{fig:jumps}.}
\bea
\label{e:flux_Q}
Q(v)=Q(v_0) + \mathrm{sign}(C)\int_{v_0}^v F_Q(v')dv' \ , 
\eea
this requiring an initial value $Q(v_0)$.
However, such coordinate $v$ is not natural in an Initial Value Problem approach.
As discussed in \ref{s:jumps}, the 3+1 slicing $\{\Sigma_t\}$ induces a 
splitting of the DH into {\em internal} and {\em external} sections.
The integration in (\ref{e:flux_Q}) can then be split into 
{\em external} and {\em internal} horizon parts (cf. Fig.~\ref{fig:inner_hor})
\bea
\label{e:news_t_truncated}
Q(t)= Q(v_0)+\mathrm{sign}(C)\int^{t}_{t_c}\left(F_Q\right)^{\mathrm{int}}(t') dt'
+ \mathrm{sign}(C) \int_{t_c}^t \left(F_Q\right)^{\mathrm{ext}}(t') dt' +
\mathrm{Res}(t) \ ,
\eea
where the error $\mathrm{Res}(t)$ is $\mathrm{Res}(t)= \mathrm{sign}(C)
\int^{\infty}_{t}\left(F_Q\right)^{\mathrm{int}}(t') dt'$.
If the growth of $Q$ is understood as ultimately 
associated with some flow into the black hole singularity,
the actual essential role of the horizon ${\cal H}$ would be that of 
capturing the associated fluxes. This assumes that the 
worldtube ${\cal H}$ {\em begins} at the {\em formation} of the singularity. 
 More complex singularity structures
(as those coming from a binary merger) would require a more detailed analysis
of this point.
From this perspective, there is nothing intrinsically special about dynamical horizons:
{\em any} hypersurface separating the black hole singularity from
past null infinity $\scri^-$ (e.g. the event horizon) would be appropriate 
for fluxes evaluation.
However, from a quasi-local perspective, if DHs are shown to {\em cover} 
systematically the black hole singularity (or, more generally, the inner Cauchy horizon),
they actually provide excellent geometric prescriptions
for such test screens (this is the motivation for the point iii) 
in section \ref{s:open_issues}).

\subsubsection{Auxiliary test-field evolutions in curved backgrounds}
In  \ref{s:recoil_crosscorrelation} we have considered cross-correlations
between different contractions of the Weyl tensor at distinct hypersurfaces.
It is legitimate to question if such cross-correlations are meaningful at all, given their
{\em a priori} different geometric contents. Let us consider the
following approach 
to this issue: evolve, together 
with the gravitational degrees of freedom in 
Einstein equations, an auxiliary (set of) scalar field(s) $\Phi_i$ 
without back-reaction on the
geometry (i.e. {\em test fields}) and whose evolution on the  
dynamically evolving
background spacetime closely tracks\footnote{See Ref. 
\refcite{Bentivegna:2008ei} for a discussion of a similar approach
in a binary black hole context, and 
Ref. \refcite{OweBriChe11} for a methodology sharing part of the spirit 
but directly tracking  spacetime curvature quantities.} its relevant  
geometric features.
Then, the correlation approach outlined in \ref{s:effective_approach} 
for a (coarse-grained) extraction of geometric content, can be applied directly on $\Phi_i$. 
We can paraphrase this approach as {\em pouring sand on a transparent surface}. 
On the one hand, this removes
the ambiguity in the choice of quantities $h_i$ and $h_o$ at inner and outer
hypersurfaces. On the other hand, and more importantly, it also permits to extend 
to the bulk spacetime the (cross-)correlation strategy between spacetime boundaries.

%%%%%%%%%%%%%%%%%%%%%%%%%%%%%%%%%%%%%%%%%%%%%%%%%%%%%%%%%%%%%%%%%%%%%%%%%%
%%%%%%%%%%%%%%%%%%%%%%%%%%%%%%%%%%%%%%%%%%%%%%%%%%%%%%%%%%%%%%%%%%%%%%%%%%
%%%%%%%%%%%%%%%%%%%%%%%%%%%%%%%%%%%%%%%%%%%%%%%%%%%%%%%%%%%%%%%%%%%%%%%%%%
\section{General perspective}
\label{s:Gen_pers}
%%%%%%%%%%%%%%%%%%%%%%%%%%%%%%%%%%%%%%%%%%%%%%%%%%%%%%%%%%%%%%%%%%%%%%%%%%
%%%%%%%%%%%%%%%%%%%%%%%%%%%%%%%%%%%%%%%%%%%%%%%%%%%%%%%%%%%%%%%%%%%%%%%%%%
%%%%%%%%%%%%%%%%%%%%%%%%%%%%%%%%%%%%%%%%%%%%%%%%%%%%%%%%%%%%%%%%%%%%%%%%%%
We have presented an introduction to some aspects of
quasi-local black holes in an Initial Value Problem approach to the spacetime
construction.
From a fundamental perspective, quasi-local black
hole horizons provide crucial insights into the geometry of the 
black hole and trapped regions and a sound avenue to black hole physics
in generic scenarios. However, quasi-local black holes also meet challenges
when considered as physical surfaces of a compact object.
We have adopted a {\em pragmatic} or effective approach in which
quasi-local black hole horizons are understood as hypersurfaces with
remarkable geometric properties that provide worldtubes of canonical surfaces
in a given 3+1 slicing of the spacetime. We have shown how they can be used
as an {\em a priori} ingredient in evolution schemes to Einstein equations,
where they provide inner boundary conditions for black hole
spacetimes. Then we have illustrated their use 
as {\em a posteriori} analysis tools tracking and characterizing
quasi-locally the black hole properties
and providing, through their {\em rigidity} properties,
excellent {\em test-screen} probes into the near-horizon black hole spacetime geometry.

\medskip

\noindent{\bf Acknowledgments.~} 
I thank the organizers of the 2011 Shanghai Asia-Pacific School,
especially C.-M. Cheng, S.A. Hayward and J. Nester, for their kind invitation 
and hospitality. I thank E. Gourgoulhon, B. Krishnan, R.P. Macedo, P. M\"osta, 
A. Nielsen and L. Rezzolla for the interactions during the writing of these notes.
I acknowledge the support of the Alexander von Humboldt Foundation, the Spanish MICINN 
(FIS2008-06078-C03-01) and the Junta de Andaluc\'\i a (FQM2288/219).

%\bibliographystyle{spmpsci}
%\bibliography{jaramillo_shanghai}

%\end{document}

\end{document}